\documentclass[a4paper,11pt]{article}

\usepackage{jinstpub} 
 \pdfoutput=1
\usepackage{lscape}
\usepackage{lineno}
\usepackage{subcaption}
\usepackage{tablefootnote}
\usepackage{hyperref} 
\usepackage{url}
\usepackage{booktabs, caption, makecell}
\usepackage{threeparttable}
\usepackage{comment}

\title{\boldmath Report of the Accelerator Frontier Topical Group 6 on Advanced Accelerator Concepts for Snowmass 2021} 

\bigskip 
\vspace{0.1cm}

\author{AF6 Conveners: \\ \textnormal{C.G.R. Geddes $^{1}$, R. Assmann $^{2}$,M. J. Hogan $^{3}$, and P. Musumeci $^{4}$}\\ Community Contributors: \\ \textnormal{E. Adli$^{5}$, C. A. Aidala $^{6}$, F. Albert$^{7}$, W. An$^{8}$, G. Andonian$^{4}$, O. Apsimon$^{10}$
% ...
}}

\bigskip 
\vspace{0.1cm}
\affiliation{
$^{1}$Lawrence Berkeley National Laboratory, Berkeley, California 94720, USA\\
$^{2}$Deutsches Elektronen-Synchrotron (DESY), Hamburg 22607, Germany \\
$^{3}$SLAC National Accelerator Laboratory, Menlo Park, California  94025, USA\\
$^{4}$University of California-Los Angeles, Los Angeles, California 90095, USA\\
$^{5}$University of Oslo, 0315 Oslo, Norway\\
$^{6}$University of Michigan, Ann Arbor, Michigan 48109, USA\\
$^{7}$Lawrence Livermore National Laboratory, Livermore, California 94550, USA\\
$^{8}$Beijing Normal University, Hai Dian Qu, Bei Jing Shi 100875, P.R. China\\
% ...
$^{10}$University of Liverpool, Liverpool L69 3BX, United Kingdom\\
% ...
$^{10}$Tsung-Dao Lee Institute (TDLI) and Shanghai Jiao Tong University, Shanghai, China\\
% ...
}

\abstract{The Snowmass 2021 Accelerator Frontier topical group \# 6 on Advanced Accelerator Concepts, covers new R\&D concepts for particle acceleration, generation, and focusing. Based on community input, this report describes how leveraging these concepts to efficiently harness the interaction of charged particles with extremely high electromagnetic fields at very high frequencies can provide the keys to reaching ultra high acceleration gradients (GeV/m and beyond). These methods have potential to reduce the dimensions, CO$_2$ footprint, and costs of future high energy physics machines, with added potential to reduce power consumption for future e+e- and $\gamma - \gamma$ machines to and beyond 15 TeV energies. Techniques range from laser and beam driven plasma and advanced structure accelerators to advanced phase space manipulations and generation of beams with extreme parameters.}


\begin{document}

\maketitle
\flushbottom

\pagebreak

\pagebreak
\section{Executive Summary}
Efficiently harnessing the interaction of charged particles with extremely high electromagnetic fields at very high frequencies is the key to reaching ultra high gradients (GeV/m and beyond) and hence to reducing the dimensions,  CO$_2$ footprint, and costs of future high energy physics machines, with the potential to reduce power consumption and offer e+e- and $\gamma - \gamma$ machines to and beyond 15 TeV energies. In addition to proven high gradient and  ultra-bright beam generation, these systems have the potential for short beams to increase luminosity per unit beam power, for practical energy recovery to extend the reach of high energy physics, and for fast cooling. Techniques range from laser and beam driven plasma and advanced structure accelerators to advanced phase space manipulations and generation of beams with extreme parameters \cite{adolphsen2022european}. Recognizing this promise the last Snowmass  and  P5/HEPAP recommended, and DOE developed with the community, an organized Advanced Accelerator Development Strategy, and work has been aligned to it  ~\cite{roadmap.doe.2016}.

In the last decade advanced accelerator research has seen tremendous progress including the demonstration of multi-GeV acceleration in a single stage \cite{gonsalves2019petawatt,Litos2015PPCF}, positron acceleration \cite{corde2015multi}, efficient loading of the structure \cite{Litos2014Nature,darcy.nature.2022}, the first staging of plasma accelerators \cite{steinke2016multistage}, demonstration of beam shaping to improve efficiency in plasmas ~\cite{rousel2020} and structures ~\cite{gao2018prl}, high gradient structures \cite{Oshea16} and greatly improved beam quality \cite{LindstromPRL2021, pompili2021energy} which recently culminated in the spectacular first demonstrations of laser-driven and beam-driven plasma based FELs \cite{wang2021,pompili2022,SOLEIL:plasmaFEL}. 

At the same time, solutions for potential collider issues have been identified, demonstrating that in principle the required nm-class emittance can be preserved (addressing potential limits due to scattering, hosing, and radiation), and that shaped bunches can be used to  efficiently accelerate beams without energy spread growth. Driver technologies (SRF linacs, high average power lasers) are developing consistent with the   luminosity and efficiency needs of future colliders. Conceptual parameter sets for colliders have been developed for e+e- and $\gamma \gamma$ colliders at a range of energies, and current evaluations indicate they present potentially competitive options~\cite{ITF22} with prospects for future cost reduction. This progress has been supported by the rapid development of high performance PIC codes, and development will continue to require the leading edge of supercomputing. 

While recent results indicate that the main building blocks of future advanced accelerators are workable and promising, significant development is still required. There are still several challenges to be addressed including how to achieve the high wall-plug efficiency and high repetition rates needed to fulfill future collider luminosity requirements, how to preserve small energy spreads and beam emittance over many acceleration stages and, for plasmas, efficient positron acceleration. 

With the goals of addressing these long-standing questions and realizing the promise of advanced accelerators, in addition to a strengthened R$\&$D program to solve outstanding critical issues, two new research directions can be identified. 
An integrated design study is needed to unite all the various elements in AAC. This process should start with the identification of an integrated set of parameters and offer a clear and actionable R\&D path to a future collider. The need is also clear to pursue nearer-term applications both inside and outside high-energy physics to increase the technology readiness level of advanced accelerators and provide a viable path to an AAC collider. 

To move forward, a vigorous R\&D program is required. The US is in a good position with  state of the art of beam test facilities focused on  advanced accelerators, including FACET-II, BELLA, ATF, AWA and FAST-IOTA, as well as numerous universities. Strong R\&D  is needed to push forward the  next steps in the Development Strategy~\cite{roadmap.doe.2016} including staging of multiple modules at multi-GeV energies, high efficiency stages, preservation of emittance for electrons and positrons, and high fidelity phase space and beam shape control, as well as  development of efficient, low-cost and high repetition rate drivers. The facilities are organized in a beam test facility council which serves to foster collaboration and minimize duplication of effort. Proposed upgrades of the  facilities (including a kBELLA high repetition rate driver/accelerator demonstrator,  positrons at FACET-II, and a GeV-class scalable module at AWA) and new R\&D are needed to maintain the US position in developing  next generation  capabilities in an environment with  $ \$ $B-class investment overseas. 

Development of a design study for compact high-gradient colliders is important to guide efforts. 
No well defined funding path 
exists for this effort, but the time is ripe to build on
the tremendous progress  over the last decade and on  recent work that has developed collider concepts and parameter sets.  
The study should provide detailed examples of how the main challenges can be addressed and clarify where experimental demonstrations, or detailed simulations, on the relevant sub-systems are needed. 
The design study should include enough detail to make cost estimates and  include strategies for demonstration colliders at moderate energies, ca. 100 GeV. 
Notably, this is in strong alignment with the European roadmap for accelerators, which includes a chapter on advanced accelerators explicitly calling for what is referred to there as an integrated design study. The European minimum requested funding for the design study is $\approx$75 FTE-years.
It is critical that to avoid losing leadership, a process occurring in many scientific areas as recently highlighted in a high-profile BESAC report \cite{besac_report}, the US should match and coordinate with this effort. 

Successful deployment of advanced concepts in real-world accelerator applications such as radiation sources for medicine, industry, security/defense, and basic science will be essential to provide the necessary intermediate steps before compact accelerators can be applied to the most demanding high-energy physics applications. The international community has long recognized the role of such near-term applications as stepping stones for high energy physics machines and has strongly invested in them (see for example the EUPRAXIA project). Even though the advanced accelerator field was born and is still squarely centered in the US, it is telling that the most recent high profile plasma-based FEL demonstrations occurred in Europe and Asia. This kind of research in the US is unfortunately not seen as directly impacting HEP, putting in jeopardy the US leadership.

At the same time, synergies with existing or near future colliders should be explored in the near term. The extremely high fields of advanced accelerator concepts could be used for transverse focusing of the beam, advanced phase space manipulations or particle sources. Possible upgrade paths of existing and near-term machines that could benefit from advanced accelerator concepts should be identified. Efforts on high brightness electron sources, polarized positron generation, high average power laser drivers, and beam delivery systems are particularly important in this regard.

While advanced accelerators in plasmas and structures continue to advance towards colliders, components, and near term applications, innovative concepts such as nanoplasmonics and laser-driven structures continue to emerge, offering the potential for even greater reach in the future. In this context, advanced accelerator R\&D and facilities serve all novel accelerator research. Furthermore, they play a critical role in accelerator and beam workforce development and diversity since they allow hands-on training with strong publication (over 1000 papers/year) for the next generation of accelerator scientists from more diverse communities who are often attracted to the field by the scientific novelty and rich physics of advanced accelerators. 

\pagebreak

Priority research should continue to address and update the Advanced Accelerator Development Strategy~\cite{roadmap.doe.2016}:
\begin{itemize}

\item Vigorous research on advanced accelerators including experimental, theoretical, and computational components, should be conducted as part of the General Accelerator R\&D program. This will advance the advanced accelerator R{\&}D roadmaps towards future high energy colliders, develop intermediate applications, and ensure international competitiveness. Priority directions include staging of multiple modules at multi-GeV, high efficiency stages, preservation of emittance for electrons and positrons, high fidelity phase space control, active feedback precision control, and shaped beams and deployment of advanced accelerator applications. 

\item A targeted R\&D program addressing high energy advanced accelerator-based colliders (e.g. to 15~TeV, with intermediate options) should develop integrated parameter sets in coordination with international efforts. This should detail components of the system and their interactions, such the injector, drivers, plasma source, beam cooling, and beam delivery system. This would set the stage for an integrated design study and a future conceptual design report, after the next Snowmass.  

\item{Research in near-term applications should be recognized as essential to, and providing leverage for, progress towards HEP colliders. The interplay and mutual interests in this area between Offices in DOE-SC, including HEP, BES, FES, and ARDAP, as well as with NSF, NNSA, Defense and other agencies should be strengthened to advance and leverage research activities aimed at real-world deployment of advanced accelerators.}

\item{Advanced accelerators should continue to play a key role in workforce development and diversity in accelerator physics. University programs and graduate students greatly benefit from the scientific visibility of the advanced accelerator field. Access to user facilities for graduate students and early career researchers as well as formal and hands-on training opportunities in advanced beam and accelerator physics should be continued and enhanced.}

\item Enhanced driver {R\&}D is needed to develop the efficient, high repetition rate, high average power laser and charged particle beam technology that will power advanced accelerators colliders and societal applications. 
 
\item Support of upgrades for Beam Test Facilities are needed to maintain progress on advanced accelerator Roadmaps. These include development of a high repetition rate facility, proposed as kBELLA, to support precision active feedback and high rate; independently controllable positrons to explore high quality acceleration, proposed at FACET-II; and implementation of a integrated SWFA demonstrator, proposed at AWA.

\item A study for a collider demonstration facility and physics experiments at an intermediate energy (c.a. 20--80~GeV) should establish a plan that would demonstrate essential technology and provide a facility for physics experiments at intermediate energy.  

\item A DOE-HEP sponsored workshop in the near term should update and formalize the U.S. advanced accelerator strategy and roadmaps including updates to the 2016 AARDS Roadmaps, and to coordinate efforts.
\end{itemize}
\pagebreak

\section{Introduction}




    

Research on advanced and novel accelerators has exploded in recent years, motivated by the potential of these techniques to increase the reach of future colliders through improved accelerating gradient, efficiency, and advanced subsystems. Progress has been driven by extremely rapid advances in performance (from the demonstration of mono-energetic 0.1 GeV beams just 18 years ago) enabled by advances  in high-power lasers based on the 2018 Nobel Prize winning technique of chirped-pulse amplification as well as ultrashort pulse shaped electron drivers (and recently high energy proton drivers). In particular, numerous high-power laser facilities have sprung up worldwide, particularly in Europe and Asia as well as high power beam test facilities. Consequently, over 1000 research papers are published annually in advanced and novel accelerators.  
This high level of activity is spawning numerous new ideas, concepts, techniques that will help bring a future advanced accelerator based collider to fruition. It also attracts students and new researchers to the field, and provides opportunities that are important for involving more diverse communities, helping to support the goals expressed in the papers ~\cite{Bai22, Arce-Lareta22, Barzi22, Hansen22} for diversity, equity and inclusion.  Since much of this research is overseas, it is critical that the U.S. make strong investments to ensure global leadership.

\begin{figure}[b]
    \centering
    \includegraphics[width=0.5\columnwidth]{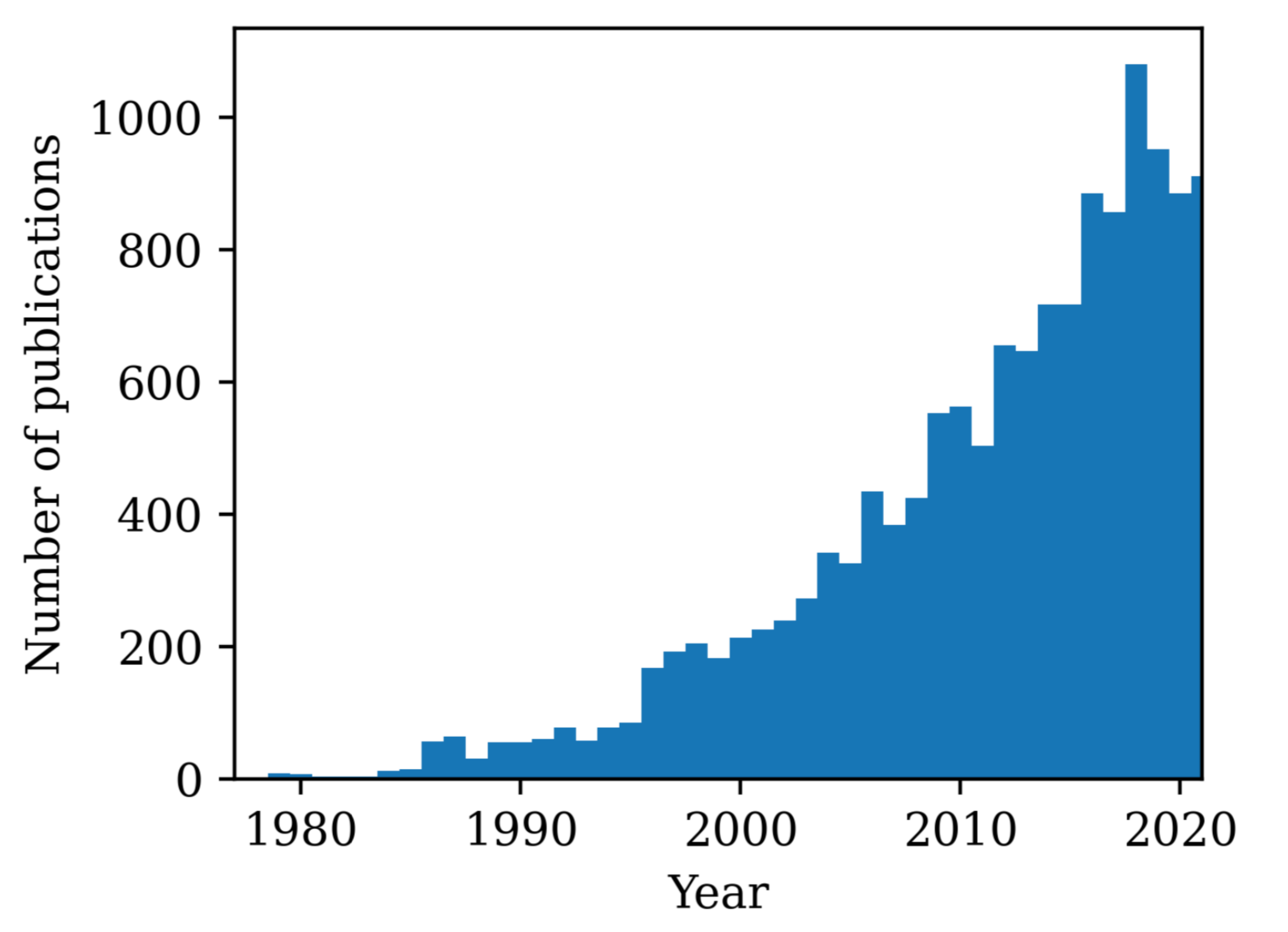}
    \caption{Example strong research, showing the number of publications per year as obtained from a Google Scholar search on articles containing all of the following keywords: ``laser$+$plasma$+$wakefield$+$accel*''.}
    \label{fig:nrofpubl}
\end{figure}

In the US, three core technologies are being pursued in a coordinated program where results from each inform the others, as organized by the 2016 DOE Advanced Accelerator Development Strategy~\cite{roadmap.doe.2016}. In a Laser Plasma Accelerator (LPA), or Laser Wakefield Accelerator (LWFA), an intense laser pulse propagating in an underdense plasma ponderomotively drives an electron plasma wave (or wakefield) \cite{Esarey09,Hooker13}. The plasma wave has a relativistic phase velocity and can support large fields suitable for charged particle acceleration and focusing. This concept is detailed in the white paper ~\cite{benedetti.arxiv.2022a}. Similar structures can be driven by the space charge force of a relativistic particle beam, known as Plasma Wakefield Acceleration (PWFA) as detailed in the white papers \cite{Gessner22, Muggli22}.  Particle beams can also drive large fields in structures (metallic,  dielectric), known as Structure Wakefield Acceleration (SWFA), which is another advanced accelerator concept as detailed in the white papers \cite{Jing22, XueyingLu22}. While in the long term these two drivers offer risk mitigation, distinct benefits (e.g. field shaping with lasers, lack of dephasing with beams), and the potential for a hybrid collider taking advantage of the best of each, in the short term the R\&D facilities explore different aspects which benefit all options. These are supported by strong progress in supporting techniques including particle sources~\cite{Fuchs22}, beam cooling, high gradient beam delivery~\cite{Gessner22b},   drivers~\cite{Kiani22}, many of which could be used for an integrated AAC collider or to upgrade a conventional collider.  Strong near term applications of advanced accelerators~\cite{Emma22, Boucher22}  reinforce the path to a collider and provide important leverage to the high energy physics program. While these technologies are progressing strongly towards collider needs as summarized in the above papers and the ITF report~\cite{ITF22}, other new techniques continue to emerge and be developed, ranging from laser driven structures~\cite{England22} to plasmonics~\cite{Sahai22} and channeling, \cite{Ariniello22} demonstrating the rich potential of future accelerators to extend the reach of physics.  

\begin{figure}
    \centering
    \includegraphics[width=0.4\columnwidth]{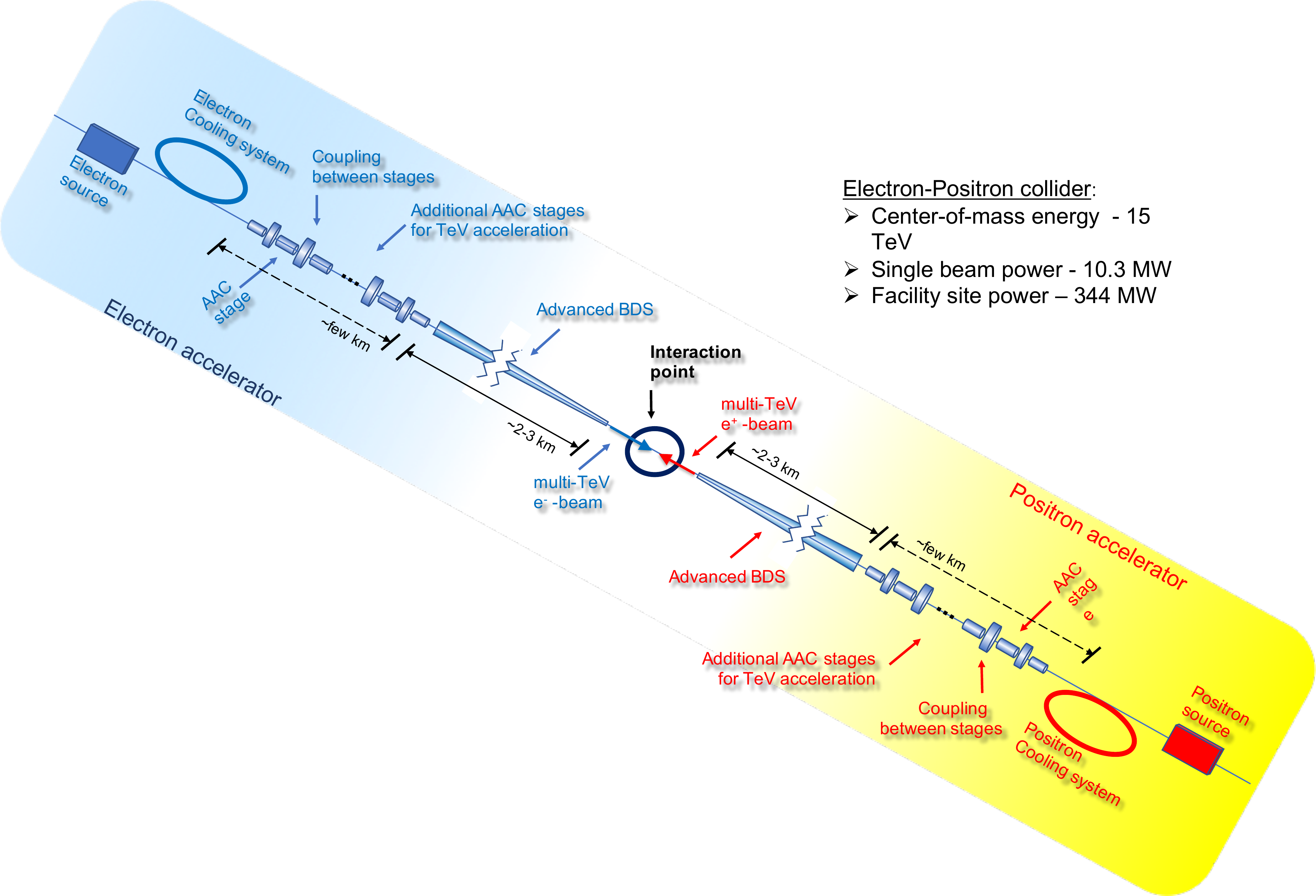}
    \includegraphics[width=0.4\columnwidth]{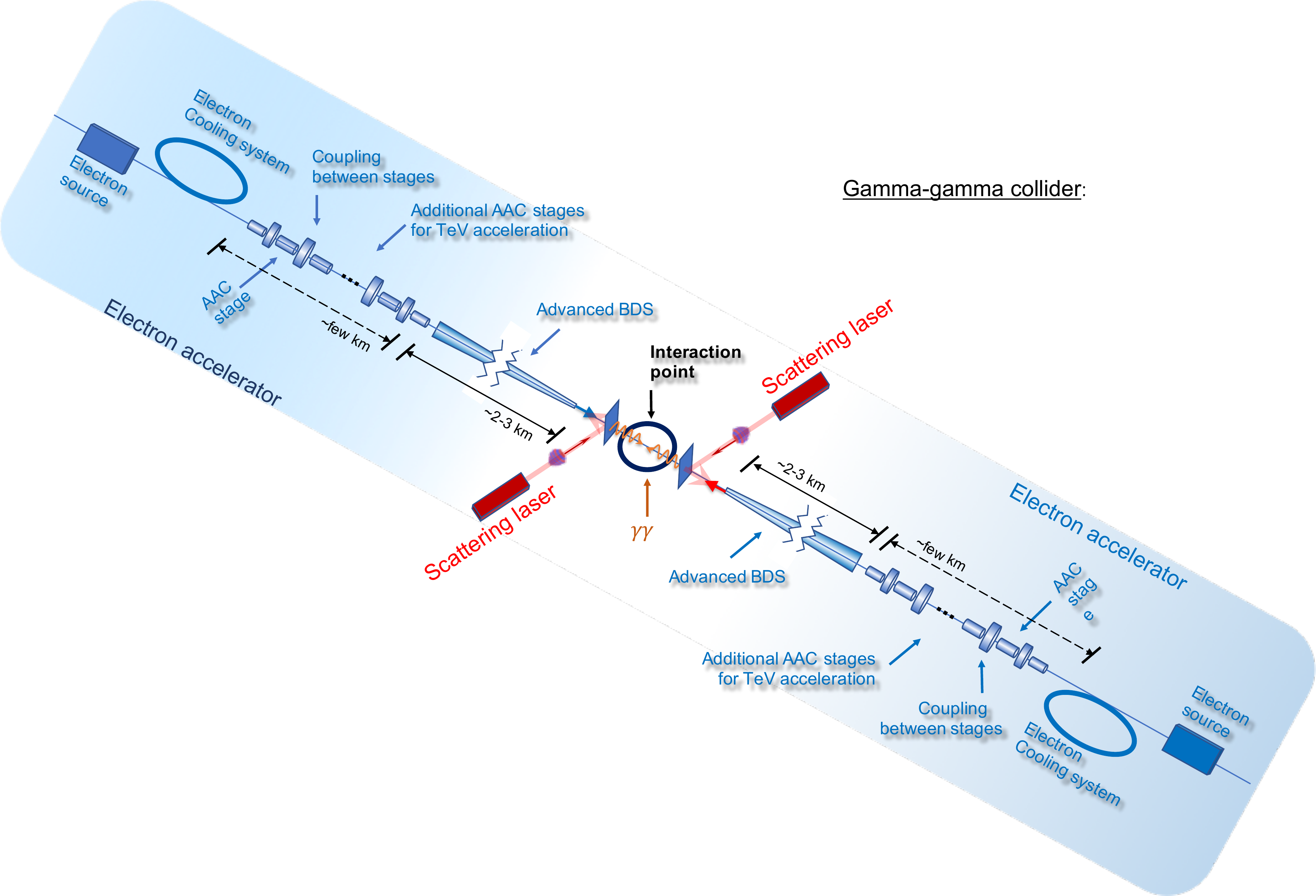}
    \caption{An advanced accelerator linear collider has the potential to support electron-positron or $\gamma\gamma$ interactions or an electron-ion collider at many TeV energies in few-km distances. Conventional or advanced injector or cooling techniques will supply beam to a series of high gradient stages, and conventional or high gradient plasma beam delivery systems.}
    \label{fig:collider2}
\end{figure}

Strategy updates developed in both Europe and the U.S. in the past decade prioritize R\&D on advanced accelerator concepts, including plasma and advanced structure based accelerators, due to their strong potential benefits. The 2014 {\it Report of the Particle Physics Project Prioritization Panel (P5)} \cite{P5-2014} highlights the importance of high-gradient acceleration technologies for future colliders (page~47):
\begin{quote}
    There is a critical need for technical breakthroughs that will yield more cost-effective accelerators. For example, ultra-high gradient accelerator techniques will require the development of power sources (RF, lasers, and electron beam drivers) compatible with high average power and high wall plug efficiency, and accelerating structures (plasmas, metallic, and dielectric) that can sustain high average power, have high damage threshold, and can be cascaded. Engagement of the national laboratories, universities, and industry will be essential for comprehensive R\&D to meet these challenges. Advancing these technologies will benefit many other areas of science and technology.
\end{quote} 
This was reinforced by a subsequent Accelerator Research and Development Subpanel {\it Accelerating Discovery: A Strategic Plan for Accelerator R\&D in the U.S.} \cite{HEPAP15}.  This report stressed the need to adequately support advanced acceleration R\&D towards future HEP colliders, as well as required demonstration facilities (pages 11 and 30). 

Similarly, the need to intensify development of high-gradient and plasma-based accelerator technology was strongly recommended in the recent report {\it 2020 Update of the European Strategy for Particle Physics} \cite{ESPPU20} (page~9):
\begin{quote}
    Innovative accelerator technology underpins the physics reach of high-energy and high-intensity colliders. It is also a powerful driver for many accelerator-based fields of science and industry. The technologies under consideration include high-field magnets, high-temperature superconductors, plasma wakefield acceleration and other high-gradient accelerating structures, bright muon beams, energy recovery linacs. The European particle physics community must intensify accelerator R\&D and sustain it with adequate resources. A roadmap should prioritize the technology, taking into account synergies with international partners and other communities such as photon and neutron sources, fusion energy and industry. Deliverables for this decade should be defined in a timely fashion and coordinated among CERN and national laboratories and institutes.
\end{quote}
This 2020 European Strategy report \cite{ESPPU20} was followed by a report of the European {\it Laboratory Directors Group} \cite{LDG22} that developed an accelerator R\&D plan and concluded (page 137): 
\begin{quote}
The field of high-gradient plasma and laser accelerators offers a prospect of facilities with significantly reduced size that may be an alternative path to TeV scale e+e- colliders. Though presently at an earlier development stage than the other accelerators, first facilities in photon and material science are now feasible and are in preparation. These accelerators also offer the prospect of near term, compact and cost-effective particle physics experiments that provide new physics possibilities supporting precision studies and the search for new particles. 
\\
The expert panel has defined a long term R\&D roadmap towards a compact collider with attractive intermediate experiments and studies. Such machines are an option for a compact collider facility beyond the timeline of an eventual FCC-hh facility. A delivery plan for the required R\&D has been developed and includes work packages, deliverables, a minimal plan, connections to ongoing projects and an aspirational plan. The panel recommends strongly that the particle physics community supports this work with increased resources in order to develop the long term future and sustainability of this field, and the US program needs similar prioritization.
\end{quote}

In addition to the HEP community, the following reports stressed their strong interest in and support of advanced accelerator technology, highlighting synergies and possible technology spin-offs including near term applications:
\begin{itemize}
    \item The 2019 report of the {\it Basic Research Needs Workshop on Compact Accelerators for Security and Medicine} \cite{BRN19} states strong interest in and support for compact 100 MeV to 100 GeV accelerators based on laser-driven wakefields (pages 44, 77, 78, 158, 193, 196, 201 and 202).
    \item The 2019 {\it Brightest Light Initiative Workshop} \cite{BrightestLightReprot18} report states that (page 3.18):
    \begin{quote}
        Extreme acceleration gradients in laser-plasma accelerators can be leveraged for future applications and light sources that need low-emittance, high-brilliance beams by investing in short-pulse laser systems with kHz to MHz repetition rates.
    \end{quote}
    \item The 2020 {\it Report of the Fusion Energy Sciences Advisory Committee} \cite{DPP20}, titled `Powering the Future: Fusion \& Plasmas' (pages 18 and 35) recommends to (page 35):
    \begin{quote}
        Pursue the development of a multi-petawatt laser facility – and a high-repetition-rate high-intensity laser facility in the US, in partnership with other federal agencies where possible.
    \end{quote}
    \item The 2021 {\it Plasma Science, Decadal Report} \cite{NAS-Decadal20} put forward five recommendations encouraging collaborative research on plasma acceleration in theory, computation, and experiments (pages 158--160).
\end{itemize}

In this report, we summarize the state of the field, drawing from the individual white papers and the ITF report to which we refer the reader for detailed material. First, we give an overview of the recent high level results that continue to advance the relevance of these accelerators to future colliders.  Next, we describe analysis of potential collider issues, which so far indicate that there are no  barriers to such a machine and motivate the development of an integrated parameters set towards an integrated design study as a next step. Context of US efforts with the recent European roadmap and international collaboration are presented.  Near term applications will use such accelerators before a collider, and we describe the important role of their development on the path towards a collider. These developments are based on a strong network of test facilities, which need continued investment to maintain progress and the US global leadership position. Other advanced concepts which continue to increase the potential of future accelerators, and the role of advanced accelerators in the accelerator community workforce are discussed.  We conclude with a summary and priority directions for the field. 

\newpage

\section{Plasma and structure based accelerator  R\&D}



    
    

The field of laser and beam driven plasma, and beam driven structure, accelerators has continued to see rapid progress over the last decade and since the last Snowmass process, addressing key components as identified in the last Snowmass process and  P5 subpanel-recommended Advanced Accelerator Development Strategy, published by DOE in 2016~\cite{roadmap.doe.2016}. These include: collider-relevant single stage energies (albeit not yet with the required emittance and charge), proof of principle staging of plasma accelerators including timing and alignment, initial capability to accelerate and focus positrons, efficient structure to beam acceleration using tailored drive beams, and greatly improved beam quality.   Driver and plasma techniques with potential to support operation at collider class rates of 10s of kHz are emerging.  As detailed in the section below, studies have also started to address combination of these elements into collider concepts addressing issues such as instability mitigation and preservation of low emittance. As described in Section 8, new and alternative concepts continue to emerge and advance the field.

Since the first demonstration of high-quality beams from LPAs in 2004 \cite{Geddes04,Faure04,Mangles04}, steady progress has been made at increasing the beam energy gain in a single plasma stage with both laser and beam drivers. The current record LPA energy gain of $7.8$ GeV was obtained in 2018 by guiding a $0.85$~PW laser in a 20 cm-long (corresponding to about $15$ diffraction lengths) laser-heated capillary discharge waveguide \cite{Gonsalves19}. Development of laser guiding techniques that allow an LPA to operate at low plasma densities, while still keeping the laser driver tightly focused over distances much longer than its characteristic diffraction length, are critical to the realization of high-energy LPAs. In this respect, several schemes for the production of meter-scale plasma waveguides using optical-field-ionization techniques have been recently proposed and validated. \cite{Shalloo19, Miao20} The next step will be to combine these with high energy drivers to push energies to and potentially beyond 10 GeV per stage and to combine this with collider relevant charge, energy spread and eventually emittance. A collider would consist of many such stages, since in plasmas the stage energy is linked to average gradient and charge. This stage energy yields minimal overall length and appropriate interaction point charge \cite{Leemans09,Schroeder10b}. 

High structure efficiency in acceleration of a discrete trailing bunch of electrons that contains sufficient charge to extract a substantial amount of energy from the high-gradient, nonlinear wake (as required for an efficient collider) has been demonstrated in a beam driven plasma accelerators~\cite{Litos2014, LindstromPRL2021}. Acceleration of about 74 picocoulombs of charge contained in the core of the trailing bunch gained about 1.6 gigaelectronvolts of energy per particle, with a final energy spread as low as 0.7 per cent (2.0 per cent on average). The experiment combined high accelerating gradient of about 4.4 gigavolts per metre and a collider relevant energy-transfer efficiency from the wake to the bunch that can exceed 30 per cent (17.7 per cent on average). New facilities using beam (FACET-II) and laser (BELLA 2nd beamline) will offer increased control of the trailing bunch to further increase efficiency.
 
High gradient structures driven by beams have been developed to GeV/m class gradients.  A high charge electron beam was used to excite a wakefield in a  dielectric-lined waveguide structure \cite{Oshea16}. Gradients of more than 1.3  GeV/m were observed in a dielectric wakefield accelerator of 15 cm length, with sub-millimetre transverse aperture, by measuring changes of the kinetic state of relativistic electron beams, and accelerating gradients of 320 MeV/m were demonstrated.  An X-band photoinjector was powered by a 400 nC drive-bunch train and  demonstrated reliable operation at 400 MeV/m with very low breakdown rates~\cite{kuzikov:ipac2021-wepab163}. These results improve on previous measurements by an order of magnitude and show promise for dielectric wakefield accelerators and X-band Photoinjectors as sources of high-energy electrons.      

\begin{figure}
    \centering
    \includegraphics[width=0.9\columnwidth]{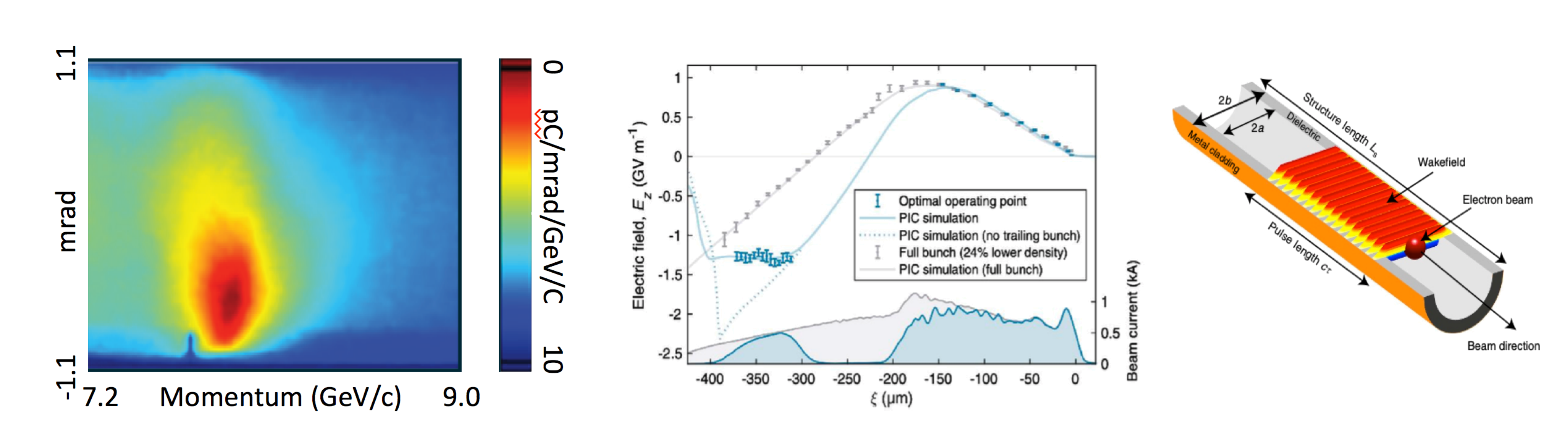}
\caption{(left) 7.8 GeV beams obtained by guiding a 0.85~PW laser in a 20 cm-long discharge waveguide (from \cite{Gonsalves19}), (center) a high transfer efficiency from structure to beam of 42\% with 0.2\% energy spread was obtained with a beam driver~(from \cite{LindstromPRL2021}); (right) GeV/m gradients were obtained in dielectric structures (from \cite{Oshea16}).}
    \label{fig:accelerators}
\end{figure}

Precise control over the 6D phase space distribution\cite{abp2021arxiv} is necessary to maximize the efficiency of an AAC-based linear collider.  Tailored distributions can also be used to mitigate collective effects and mitigate the beam breakup instability. An emittance exchange beamline was recently used to shape the longitudinal current profile of a high-charge electron bunch to demonstrate record high transformer ratio.  Such a beam was used to excite nonlinear plasma wakefields in a hollow-cathode plasma cell and a transformer ratio of R=7.8 was measured~\cite{rousel2020}.  Using the same emittance exchanger, researchers measured a transformer ratio of 5 in a collinear dielectric wakefield structure\cite{gao2018prl}. These results are key steps in advancing PWFA and SWFA technology on the roadmap towards a viable linear collider.

A proof-of-principle demonstration of staging was recently made using two independent laser-driven plasma stages \cite{SteinkeStaging}, establishing that timing and alignment are possible. In this proof-of-principle experiment, stable $\sim 100$ MeV electron beams from the first LPA (length $\lesssim 1$ mm) were focused by a discharge capillary-based active plasma lens of length 1.5 cm into a 3.3 cm-long LPA powered by a second laser pulse that was in-coupled into the second stage by means of a plasma mirror. The second LPA stage operated in a dark-current-free, quasi-linear regime, and captured and accelerated by $\sim 100$ MeV a fraction of the beam.  Future experiments are being prepared to explore multi-GeV staging and high capture efficiency including the newly commissioned BELLA 2nd beamline and others.  In a similar effort, a proof-of-principle SWFA experiment yielded the first experimental demonstration of a staged two beam wakefield accelerator\cite{JING201872} at the AWA, where a high-brightness 0.5 nC electron bunch gained equal amounts of energy in two stages corresponding to an average acceleration gradient of 70 MeV/m.  Relevant physics of stage in and out coupling and emittance preservation will also be studied at facilities such as FACET-II, ATF and others. 

Acceleration of positron beams was demonstrated using a beam driven plasma wakefield, a important step towards a future collider. A single, high energy (20 GeV), high charge ($\sim$1 nC), short ($\sim$30 $\mu$m) positron bunch was launched into a 30 cm long lithium heat pipe oven plasma source. The results surprisingly produced a monoenergetic positron beam that gained several GeV of energy, revealing a previously unpredicted regime of nonlinear positron PWFA physics ~\cite{Corde2015}. Positron wake generation and loading were also explored in experiments using hollow channels created by a Bessel laser beam  ionizing a tube of neutral gas~\cite{Gessner2016}, a promising approach for future colliders\cite{Schroeder16} that enable symmetric acceleration of electron and positron beams. 

\begin{figure}
    \centering
    \includegraphics[width=0.7\columnwidth]{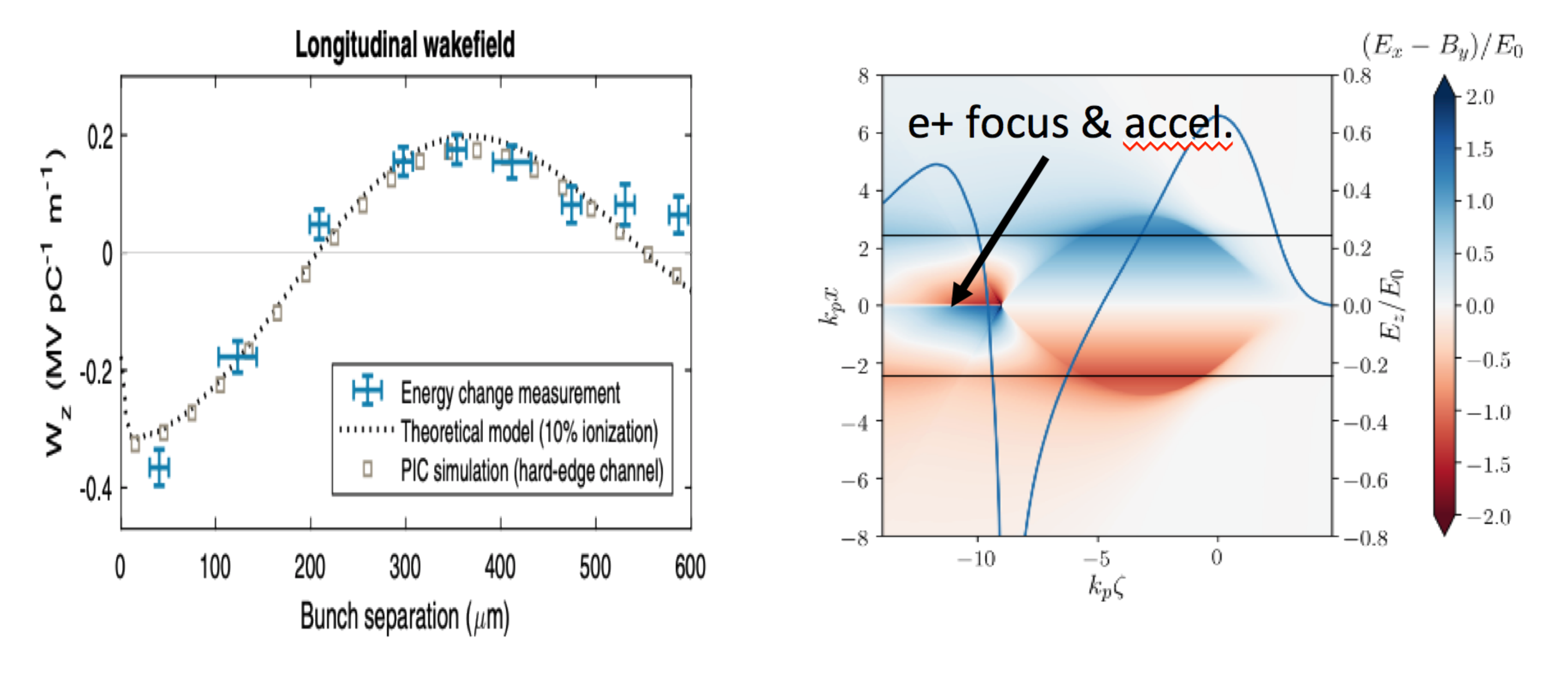}
    \caption{((left) Hollow channel experiments with a positron probe beam established a high gradient positron structure (from \cite{Lindstrom18posi}), (right) simulations indicate that new concepts including plasma columns from (\cite{Diederichs2019}) have the potential to combine high gradient and low emittance positron acceleration}
    \label{fig:positrons}
\end{figure}

The Advanced Proton Driven Plasma Wakefield Acceleration Experiment (AWAKE) experiment demonstrated the ability to accelerate electrons in a high-energy proton beam-driven plasma wakefield accelerator (PDWFA) ~\cite{Adli2018}, detailed in~\cite{Muggli22}. This takes advantage of existing infrastructure to produce high energy, high charge proton beams and to transfer that energy to an electron beam over a relatively short distance.

Plasma lenses can focus electron beams with strengths several orders of magnitude stronger than quadrupole magnets. In passive plasma lenses~\cite{Doss2019,chen:1989prd,chen:1990prl}, the transverse focusing force in the underdense, nonlinear blowout plasma wake regime is due to the presence of the stationary plasma ions. Active plasma lenses~\cite{vanTilborg15} can focus particles of either charge using an axial current carried in a plasma column, which creates an azimuthal focusing field, and focusing of high brightness beams~\cite{Pompili18} and emittance preservation~\cite{Lindstrom18} have been demonstrated. Both are far stronger than conventional quadrupoles. Such lenses were a key part of the recent staging demonstration~\cite{SteinkeStaging}, allowing coupling of stages at cm scale, and emittance preservation  ~\cite{Lindstrm2018} has been demonstrated. These plasma lenses may provide both compact coupling between stages to preserve gradient, and  additional options for beam delivery subsystems.  Continued R\&D is needed to establish  methods for collider scale beams.


Particle sources based on plasma accelerators could enable beams with tens of nm-rad normalized emittances due to the very high fields these structures sustain, and regimes with transverse field linearity, as detailed in~\cite{Fuchs22}. Ultralow-emittance beams have been demonstrated using plasma density downramps to control injection trajectories, ~\cite{Bulanov98, Geddes08, Gonsalves11,Goetzfried20, Deng2019,KnetschTorchPRAB2020} which  may give rise to slice-normalized emittances down to tens of nm rad \cite{XuPRABdownramp2017}. It offers huge tunability  with regard to the downramp profile  in 3d-space and time, \cite{UllmannPhysRevResearch2021} which experiments using laser- preformation~\cite{Wittig2015Downramp} and hybrid LWFA-PWFA \cite{CouperusDownrampPRR2021} are developing. Ionization can birth particles matched in the structure for nm-class beams.  Multiple colliding laser pulses can offer advanced control of phase space~\cite{Esarey97, Faure07, Rechatin09}.  Advances in injection supported greatly improved beam quality which recently culminated in the spectacular first demonstrations of laser-driven and beam-driven plasma based FELs \cite{wang2021,pompili2022}.  So far LPAs have demonstrated the production of high-quality electron beams with individual HEP-relevant parameters, such as relative energy spreads to and below $1\%$ \cite{Rechatin09,  Geddes16, Kirchen21,  Ke2021}, normalized emittances of $\sim$ 0.1~$\mu$m \cite{Plateau12, Weingartner12}, durations inferred to be 1-3 fs~\cite{Lundh11,Buck11}, and high charge (100s of pC) \cite{cuperus17,Goetzfried20}, even though the best parameters were not all achieved, in general, simultaneously. Simulations indicate reaching these parameters simultaneously is feasible with plasma controlled injection. This  has the potential to circumvent the need for the damping rings required in conventional accelerators, which are not capable of directly producing beams with emittances that allow for the required luminosities. There is the potential also for directly generating polarized beams~\cite{Wu19, Wen19} and for rapid linear cooling systems based on plasma undulators  \cite{Rykovanov15,Wang17}. 

Reaching the necessary luminosities for a linear collider will likely require the acceleration of thousands of bunches per second.
Driver technologies (SRF linacs, high average power lasers) have developed substantially in the last 10 years. For laser driven approaches, new diode-pumped solid-state laser technologies are needed to address collier rates and efficiency as detailed in~\cite{Kiani22}.  The fiber laser solution, combining many ultra-short pulses generated from fiber amplifiers in time, space, and wavelength \cite{Zhou:15,Zhou:18,Chang:13}, offers high wall-plug efficiencies and excellent thermal management. Hundred-mJ-class sub-scale demonstrations integrating representative components are in progress at collider relevant rates. A flexible high-rate laser driver utilizing bulk Tm:YLF crystals \cite{Galvin19,Siders19} near 2~$\mu$m wavelength is an inherently high-efficiency, single laser beam solution and recently joule-class extraction (at low average power to date) has been demonstrated. Cryogenic Yb is a high maturity option operating at joules at kHz. This indicates that multiple feasible driver paths exist, and sustained R\&D will be needed to reach technical readiness for application as LPA collider drivers \cite{Workshop13,Workshop17,falcone20}.
SRF  technologies for beam drivers have similar maturity to other accelerators to meet needs of future colliders, and require specific bunch formats. The recovery time of a beam-driven plasma accelerator was recently measured for the first time at FLASHForward at 63 ns.

\begin{figure*}[htbp]
\centering 
\includegraphics[width=0.8\textwidth]{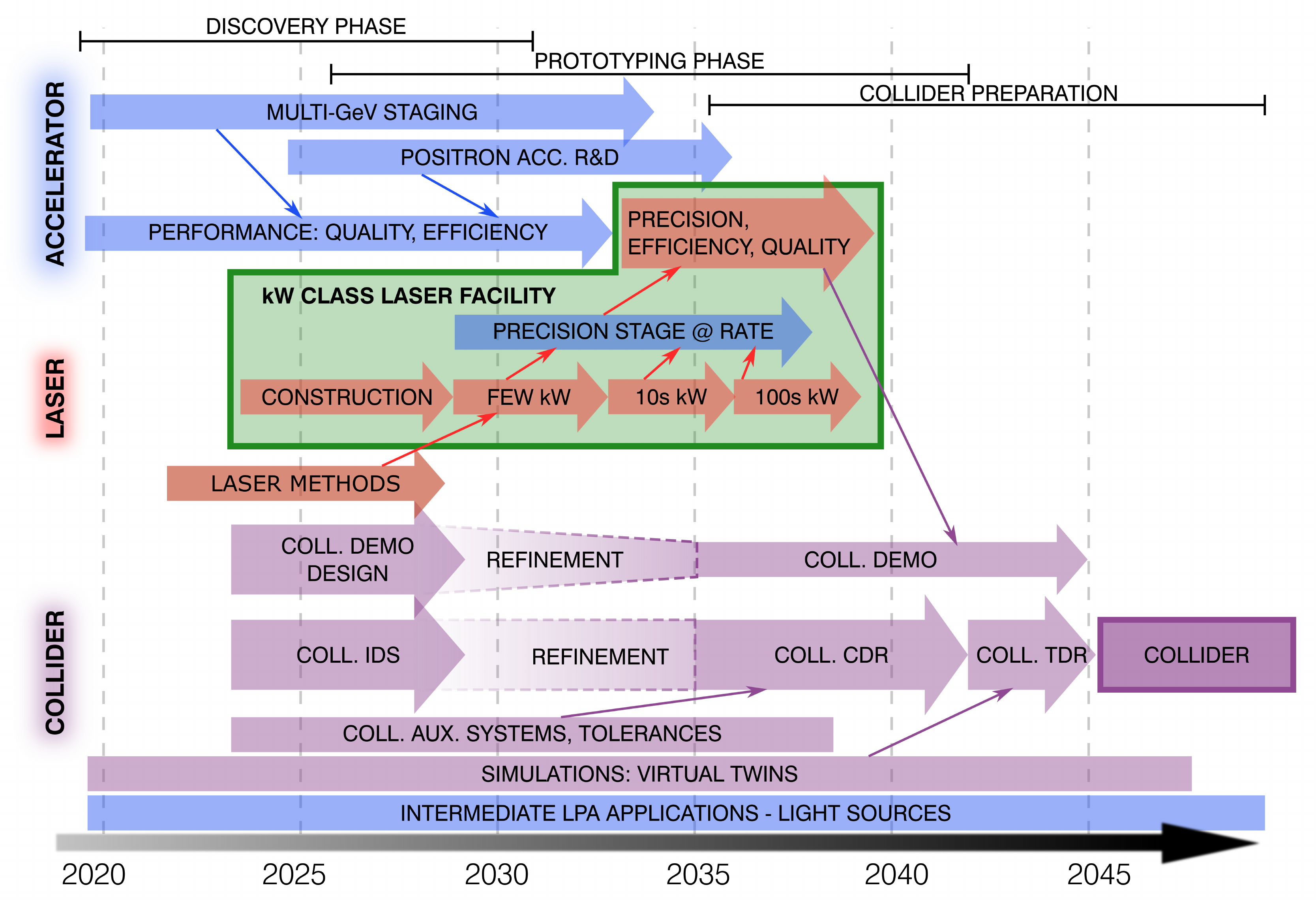}
\caption{\label{fig:new_road} Example technically-limited high-level R\&D roadmap, for laser-plasma-accelerator-based colliders. While details vary, similar roadmaps apply to other technologies as detailed in ~\cite{DOE2016} and in ~\cite{benedetti.arxiv.2022a, benedetti.arxiv.2022b, Gessner22, Jing22}}. 
\end{figure*}

Progress over the past decade has indicated that the main building blocks of a future collider are potentially workable, have the potential to offer significant scale and cost advantages, and should be developed fully. Key issues identified by the past Snowmass, P5, Accelerator R\&D subpanel, and subsequent Advanced Accelerator Developement Strategy have been addressed to the extent realistic within funding limits. These include experimental demonstration of multi-GeV stages, efficient acceleration, and initial proofs of principle of staging of multiple modules and positron acceleration. The development of large scale simulations supported these results and indicate they can be scaled to collider relevant performance. The community is positioned to take the next steps in development, including:
\begin{itemize}
\item Development of high energy electron stages towards collider charge and efficiency, using techniques such as matched laser guiding and shaped electron drive beams enabled by new facilities
\item Higher energy staging with high capture efficiency and techniques for preservation of emittance
\item Development of sub-0.1 micron emittances and their transport using advanced optical injection mechanisms and plasma shaping techniques
\item High quality positron acceleration, accessible using extensions of FACET-II and  BELLA
\item Experiments at high repetition rate, accessible on new facilities as detailed below
\item Development of supporting technologies such as efficient high repetition rate drivers, plasma lenses for high gradient focusing, diagnostics and advanced simulations
\end{itemize}
A typical technically limited roadmap for the R\&D is outlined in Figure~\ref{fig:new_road}. As detailed in the following section, conceptual parameters have been developed for future colliders. As detailed in section 7 experimental test capabilities have been developed for staging (BELLA 2nd Beamline), together with capabilities for high performance PWFA with multiple bunches (FACET-II) and structure wakefields (AWA and FACET-II). Capabilities for positron tests (FACET-II Positrons) and kHz experiments (kBELLA) have been  proposed. 

\pagebreak

\section{Collider Concepts and Integrated Design Study}

Advanced wakefield based accelerator technologies provide a path to a multi-TeV lepton collider. The high-gradient structures significantly reduces the length of the accelerating sections. The  wakefield accelerators naturally accommodate short bunches, which reduce beamstrahlung and enhance luminosity. Low emittance beams can also be generated in plasmas, which also increases luminosity. Wakefield accelerators can also support efficient transfer of energy from the drive bunch to the witness bunch. These factors yield a novel, compact, less-expensive accelerator option for exploring energy-frontier physics. The advanced accelerator based collider concepts can be considered as a stand-alone concept or as an upgrade to an existing sub-TeV linear collider facility.

Detailed conceptual designs of AAC-based colliders have not been completed. However, several preliminary studies have been performed (for details, see white papers submitted to Snowmass21 proceedings ~\cite{benedetti.arxiv.2022a,benedetti.arxiv.2022b, Gessner22,Jing22}).  These studies have been used to help identify the expected operational  parameters to guide the R\&D toward developing AAC-based collider concepts (see, e.g., Ref.~\cite{ANAR17}).  The general approach is based on wakefield accelerator modules and staging. Each wakefield module is powered by a separate drive beam (electron or laser). Depletion of the drive beam energy by exciting wakefields within the module ultimately limits the length of an individual modules, although there may be other considerations that determine the optimal length. Each module would yield an energy gain of 1--10~GeV, depending on the specific design and approach. Higher energies are obtained by coupling, or staging, additional modules. Typical summary parameters are presented in  Table~\ref{tab:collider} and in ~\cite{ITF22}.

\begin{table}[b]
\caption{\label{tab:collider} High-level collider parameters}
\begin{tabular}{lccc}
Center-of-mass energy [TeV] & 1 & 3 & 15 \\
Beam energy [TeV] & 0.5 & 1.5 & 7.5\\
Luminosity [$10^{34}$ cm$^{-2}$ s$^{-1}$] & 1 & 10 & 50\\
Particles/bunch [$10^9$] & 1.2 & 1.2 & 1.2\\
Beam power [MW] & 4.4 & 13 & 65 \\
RMS bunch length [$\mu$m] & 8.5 & 8.5 & 8.5 \\
Repetition rate [kHz] & 47 & 47 & 47\\
Time between collisions [$\mu$s] & 21 & 21 & 21 \\ 
Beam size at IP, x/y [nm] & 50/1 & 10/0.5 & 4/0.25\\
Linac length [km] & 0.22 & 0.65 & 3.3 \\
Facility site power (2 linacs) [MW] & 105 & 315 & 1100 \\
\end{tabular}
\end{table}

In the past decade, there have been significant advances in addressing collider issues. Theoretical studies have demonstrated that in principle the required nm-class beam quality can be preserved in future colliders (addressing potential limits due to scattering \cite{KirbyPAC07,zhao2020modeling}, hosing \cite{Huang07,mehrling.prl.2017}, matching\cite{WLu.prl.2006,XXu.prl.2016,ariniello.prab.2019}, and ion motion \cite{An17,benedetti.prab.2017,Mehrling2018,burov.arxiv.2018}), and that shaped bunches can be used to efficiently accelerate beams without energy spread growth \cite{gao2018prl,rousel2020}. The high gradients accessible may make beam and/or laser energy recovery practical for improved efficiency. These indicate that fundamentally an AAC collider is likely to be able to access many-TeV energies, and form the basis for future integrated designs.

It is envisioned that an AAC-based high-energy collider would initially operate at TeV center-of-mass energies, and be upgraded to $>$10~TeV. Parameters for center-of-mass energies of 1~TeV, 3~TeV, and 15~TeV for the various AAC technologies have been studied. Further discussion of these parameters and details can be found in the individual white papers. Note that the 1~TeV and 3~TeV colliders are envisioned as precision machines and use flat beams (asymmetric emittances in vertical and horizontal planes) to reduce the beamstrahlung background.  The   15~TeV  collider is envisioned as an energy frontier machine and uses round beams to reach the high luminosity at reduced power. It should be emphasized that these numerical examples are representative of what could be achieved using AAC-based linacs, and optimization will be performed during the development of a future integrated design study. AAC technology is rapidly advancing, with new innovations appearing at a swift pace, and we fully expect that the AAC collider designs and numerical parameters will evolve as new innovations and inventions in this field emerge.

Owing to the ultra-high accelerating gradients, to reach several TeV beam energies, an AAC-based linac would be of km scale.  For example, a geometric gradient of 2 GV/m (which is conservative), requires only 0.5 km in each linac arm to reach 1 TeV beam energies.  This offers the possibility of re-using the tunnels of an existing conventional linear collider complex to greatly extend the energy reach of the facility as a future upgrade. For example, one could consider using the ILC infrastructure (or some other RF-based linear collider) to upgrade the beam energy by replacing the  RF cavities with AAC stages. Upgrading to $\sqrt{s}=$1 TeV or 3 TeV, with the parameters in \cite{Gessner22,benedetti.arxiv.2022a,Snowmass:ILC} could be supported by the ILC site power. The unused main linac tunnel length could be employed to extend the Beam Delivery System to accommodate the higher beam energies, as well as space for linear cooling sections to further reduce the beam emittance. Additional bunch compressors would be required to achieve the short (10-$\mu$m-scale) bunch length needed for  high frequency wakefield accelerators. Furthermore, achieving high beam energies is straightforward by adding additional AAC stages. The achievable luminosity would be limited by the site power of the repurposed ILC design. Hence, AAC technology may provide a long-term upgrade path to continue realizing new physics reach in realistic stages using the infrastructure of a RF linear collider facility.

Accelerators are essentially a power source and a structure. In high-gradient advanced accelerator concepts the structures can be made of metals, dielectrics or plasmas and the power source is typically a high-current electron beam or high-power laser pulse. The advanced accelerator based collider concepts discussed in this section align with the three directions supported by DOE HEP General Accelerator R\&D (GARD) and combine structures and power sources as follows: (1) laser-driven plasma-based accelerators, also referred to as laser-plasma accelerators (LPAs) or as laser wakefield accelerators, (2) beam-driven plasma-based accelerators, also referred to as plasma wakefield accelerators (PWFA), and (3) beam-driven structure-based accelerators, also referred to as advanced structure wakefield accelerators (SWFA). For details on these approaches, as well as numerous references, please see the individual white papers on these topics \cite{benedetti.arxiv.2022a,Gessner22,Jing22}. Rapid progress has been made on individual accelerator components as highlighted in Section 3. The international landscape is described in Section 5 and the importance of developing and demonstrating near term applications as stepping stones towards deployment of AAC technologies is discussed in Section 6. In the paragraphs below we highlight high-level priority research directions for the different technologies and in Section 7 describe the accelerator test facilities and upgrades envisioned to enable this research program.

For Laser Plasma Accelerators (LPA) research is focused on both the accelerator technology and the auxiliary systems compatible with laser-plasma-accelerated particle beams.  Multi-GeV staging is the highest near-term research priority for LPA development, including the development of compact methods to couple the particle beam between stages, minimizing beam loss and preserving beam emittance. Advanced laser-plasma injection techniques \cite{Fuchs22} should be pursued to generate ultra-low emittance (<100 nm) electron beams that will have applications in addition to colliders while also serving as tools to develop techniques to measure and transport the low emittance beams needed for collider applications. Current high-peak-power laser systems operate at a few Hz repetition rates, and laser-plasma acceleration techniques, including plasma targetry, laser shaping, diagnostics, etc., will need to be developed for high repetition rates (tens of kHz). Laser technology is advancing rapidly, driven by applications ranging from industry to defense. Recent studies have identified how these advances can be leveraged to create plasma accelerators driven by high peak and high average power short pulse lasers \cite{Kiani22}. Demonstration facilities and experiments are envisioned as next steps for developing suitable driver technology (see Section 7). 

In a beam-driven plasma accelerator, the plasma is a transformer that takes a high-current low-voltage electron beam and converts it to a high-voltage lower-current beam in a series of compact plasma cells. For example, a 10 GeV beam with a 1 MHz repetition rate, similar to systems now being deployed for XFELs around the world, could be transformed into a 1 TeV beam with a repetition rate of 10 kHz for collider applications. The heart of such a system is the plasma cell. and recent demonstrations have operated with gradients and efficiencies proposed for plasma collider concepts. The challenge now is to achieve the same high-gradient and high-efficiency acceleration already demonstrated and described in Section 3 while also preserving the beam emittance. As in laser plasma accelerators, this involves properly matching the beams into the strong focusing within the plasma, controlling chromaticity as well as mitigating the beam break-up instability, ion motion and other mechanisms of emittance growth. External injection of beams into plasma accelerator stages can at best preserve the beam quality of the incoming beam. Internal injection methods which may produce beams with tens of nm rad normalized emittances are being developed as surrogates to study collider quality beams. Staging not only enables reaching high particle energies, but can also be used for passive feedback and correction systems that can greatly improve the beam quality and stability \cite{LindstromArxiv2021}. To maintain the high average accelerating gradient, compact methods for interstage geometries need to be developed that scale well to very high energy. Ideas exist involving plasma lenses but limitations may arise at high intensities and need to be investigated. Current state-of-the-art plasma-accelerator experiments typically operate at the hertz level. 

The inter-bunch frequency in a plasma-based collider will be defined by the time it takes for the plasma to recover to approximately its initial state after the long-term plasma motion caused by driving a wake and significant developments in plasma-source technology will be required. Although there are differences between the way laser and electron beam drivers propagate in a plasma, much of the physics for accelerating beams to high energy with low emittance is common for both approaches.  Designs will need to account for precision alignment and jitter tolerances, and high efficiency coupling of many stages, in small accelerator structures. Oftentimes advances in theory, modeling and experimental techniques from one field benefits the other and there is a strong synergy between the goals of each technology.

It is worth noting that an alternative form of beam driven plasma acceleration using high-energy proton beams as plasma wakefield drivers is being investigated by the AWAKE Collaboration, centered in Europe \cite{muggli.awake.snowmass}. This technique takes advantage of the availability of high-energy, relativistic proton bunches (such as those available from proton synchrotrons) to drive large amplitude wakefields (GV/m) in a single plasma over distances sufficient to produce hundreds of GeV to TeV electron bunches. The relatively low repetition rate of the proton bunches has led to consideration of applications other than a traditional electron-positron collider, such as replacing secondary particle sources for dark photon search experiments or TeV electron-proton collisions for very-high energy electron-proton or electron-ion collider. 

Structure Wakefield Accelerators come in two formats, the short-pulse Two-Beam Accelerator (TBA) and the Collinear Wakefield Accelerator (CWA)\cite{gai2012short}. They are distinguished by whether the drive and main beams traverse the same structure. In the short-pulse TBA configuration, the drive beam energy is extracted and converted into an RF pulse in a power extraction and transfer structure, that is then used to accelerate the main beam in a separate accelerating structure. In both cases, emittance preservation can be obtained through shaping of the electron beam and designing and testing structures that mitigate wakefields and damp higher order modes \cite{Simakov_2022}. For the TBA approach, RF breakdown is mitigated by operating with high-gradients (>300 MeV/m) and RF pulses an order of magnitude shorter than conventional techniques (10 ns). In the CWA approach, producing high-charge beams with shaped longitudinal profiles increases the gradient and energy boost per stage,  and robust techniques for shaping these high-current beams are under development. High gradients and high transformer ratios have been demonstrated and programs \cite{zholents:ipac2020-tuvir08} will scale up the total energy gain to more than 100MeV while preserving the accelerated beam emittance and then focus on staging two or more cells together for more than a GeV energy gain. Controlling wakefields, limiting emittance growth and maximizing efficiency with shaped beams are common elements to all three AAC schemes, leading to synergy in research activities and a collaborative approach across technologies. 

With core research on high gradient accelerator stages showing strong progress towards collider needs, innovative ideas are emerging for supporting systems such as the beam delivery system (BDS). The BDS transports the beam from the accelerator and brings it to a focus at the Interaction Point and includes diagnostic sections for measuring the beam energy, emittance, and polarization, as well as collimators for machine protection. The length of the BDS system increases with collision energy and in the absence of new ideas could dominate size for advanced accelerators. Such novel ideas for a future BDS systems that will reduce the size of the system and increase the luminostity of linear colliders based on Advanced Accelerator techniques are detailed in~\cite{Gessner22b}. Plasma lenses can reduce chromaticity, increase energy acceptance, and reduce the length of the final focusing elements, while novel collimation techniques can reduce length. Luminosity enhancement can be realized due to naturally short beams from Advanced Accelerators, and may be complemented using round beams or novel Machine Detector Interface (MDI) designs that can accommodate plasma lenses and large backgrounds.

Development of enhanced modeling capabilities is required in order to guide the R\&D effort towards a TeV collider. For example, the numerical modeling, in 3D, of plasma accelerators using conventional Particle-In-Cell (PIC) codes is a computationally challenging task. Modeling a collider requires simulating a chain of hundreds of stages together with all the optics required for inter-stage beam transport and driver in-coupling. In addition, ensemble runs of simulations on large parameter space will be required to estimate tolerances and assess the impact of non-ideal effects (e.g., misalignment, tilts, etc.). The development of simulation tools needed for the design of a multi-TeV collider will require robust and sustained team efforts based on collaborations in the accelerator modeling community, as well as coordination between national laboratories and university groups. These collaborations should leverage the past, ongoing, and future efforts from the DOE SciDAC program and the Exascale Computing Project.

Advocates for linear colliders are primarily interested in electron-positron collisions. This is because electron-positron collisions are ``clean": collisions between electrons and positrons have a well-defined center of mass (CM) and zero initial quantum number. In addition, the CM energy and beam polarization can be scanned to precisely probe physics near a particle resonance. Electron-positron collisions require high-energy positron beams, and this presents a unique challenge for the plasma based concepts because the acceleration of positron beams in nonlinear plasma wakefields is different from the electron beam case. As described in Section 3, many alternatives to the non-linear blowout regime have been proposed and tested yielding many of the qualities needed for a collider: multi-GeV/m accelerating gradient, multi-GeV energy gain and percent level energy spread. For collider applications, further advances are needed that deliver all of the key ingredients at the same time: high-gradient acceleration, multi-GeV energy gain, high efficiency and emittance preservation. A variety of ideas have been put forward that propose engineering the drive beam \cite{vieira2014nonlinear,zhou.e+.prl.2021} and/or the plasma source \cite{Diederichs2019,Diederichs20,silva2021stable} to meet these demands and and will be testable with facility upgrades discussed in Section 7. Additionally, plasma accelerators offer techniques to reduce the cost and footprint of the positron source, including bremsstrahlung radiators or undulators driven by compact accelerators, and the potential for compact linear cooling mechanisms such as plasma undulators \cite{Rykovanov15,Wang17}. These could benefit any future collider. 

An alternative that does not require positron acceleration is to construct a $\gamma$-$\gamma$ collider, that is, an $e^-e^-$ collider with conversion of the $e^-$ beams to $\gamma$s near the interaction point, which is one of the examples in \cite{benedetti.arxiv.2022a}.  This conversion can be done by Compton scattering from a low-photon-energy laser beam~\cite{1998hep.ex10019T}.  A scheme using highly compressed e- bunches is also being studied~\cite{Yakimenko2019}.  A  $\gamma$-$\gamma$ collider at the Higgs boson resonance would be interesting in the near term; this would require only 80 GeV/e- beam, assuming 80\% energy transfer to the $\gamma$s.  For longer-range goals, a $\gamma$-$\gamma$ collider may be just as effective as an $e^+e^-$ collider in exploring new particles with masses in the energy region of 10 TeV and above \cite{peskin.july.2022}.

A critical advance will be the development of an integrated parameters set, towards an integrated design study, for an AAC-based collider that will include preliminary designs for all the collider subsystems. Key components include the electron source, positron source, cooling system, bunch compressors, the accelerator, and the beam delivery system (BDS) to the interaction point (IP). This should be an international effort, seeking to enhance AAC collider designs by leveraging the decades of experience of conventional collider development and drawing from the best elements of each AAC technology. Beyond the next decade, it is envisioned that a low-energy (few tens of GeV) collider demonstrator would be a key step and may also provide physics opportunities~\cite{benedetti.arxiv.2022b} including (in commonality with muon colliders) potential options  for electroweak
 precision measurements~\cite{Blondel98}.  Given the cost of such a facility, a robust science case  needs to be developed (e.g., such an energy range may have potential applications to nuclear physics and QCD experiments).  With sufficient funding, such a facility could be constructed and commissioned on the 20-year timescale.   It is anticipated that, with sustained funding, a technical design report for a multi-TeV collider could be completed in the 25+ year timescale. These are  captured as high-level recommendations within the Executive Summary in Section 1. 

\newpage
\section{European Roadmap and international collaboration}

Europe has established a strong R\&D program in advanced accelerators. Several European projects and plans have been presented in Snowmass meetings or in white papers. In Europe, future R\&D plans and goals for particle physics machines have recently been assessed by an expert panel and the findings were published in the European roadmap report for accelerator R\&D; see \cite{EuroRoadmap2022}. 

The advanced accelerator R\&D in Europe is driven forward in an integrated effort, covering usage of this new technology for the full range of beam energies relevant for applied and fundamental research (both HEP and other applications). At present, first user facilities involving plasma accelerators are being set up for photon and material science in the European research landscape. The "European Plasma Accelerator with eXcellence In Application" (EuPRAXIA)\cite{Assman2020} is an officially selected and government supported project on the roadmap of the "European Strategy Forum on Research Infrastructures" (ESFRI), aimed at 5 GeV electron beams of high quality. National and European funding support for EuPRAXIA has reached 150 M€ and an EU-funded Preparatory Phase project is preparing its full implementation. Latest R\&D achievements in Europe include the Nature-published free electron lasing of an electron beam after passing through a beam-driven plasma accelerator, demonstrated at the EuPRAXIA SPARC\_lab facility in LNF/INFN\cite{pompili2022}. 

As described in the report of the European expert panel, national and regional activities in Europe will continue through the end of the 2020s with a strong R\&D and construction programme, aiming mainly at low energy research infrastructures, for example to drive a free-electron laser or to deliver ultrafast electron diffraction. Various important milestones have been and will be achieved in internationally leading programmes at CERN, CLARA, CNRS, DESY, various centres and institutes in the Helmholtz Association, INFN, LBNL, RAL, SCAPA, and others. New European research infrastructures (RI) involving lasers and plasma accelerator technology have been driven forward in recent years, namely ELI and EuPRAXIA, both placed on the ESFRI roadmap. 

The European expert panel found that plasma-based accelerators have produced multi-GeV bunches with parameters approaching those suitable for a linear collider. It concludes that significant reduction in size and, perhaps, cost of future accelerators can therefore in principle be envisaged, including  benefits for particle physics machines. The European expert panel recommends in its report that the work directed at applied research machines should be complemented by early tests and R\&D activities targeted at High Energy Physics (HEP). Given that funding for ongoing activities in Europe is mostly from non-HEP sources, several HEP-related aspects are currently not prioritised, for example: staging to high energy; efficiency; acceleration of positron bunches; and beam polarisation. It is however noted that the AWAKE collaboration and experiment at CERN is fully funded by Particle Physics and has developed the usage of proton bunches and their high stored energy as drivers of plasma wakefields to accelerate electrons \cite{AWAKE-Nature}. AWAKE has a clear science roadmap for the next decade, which aims to bring the proton driven plasma wakefield acceleration technology to a stage where first particle physics experiments can be proposed~\cite{Muggli22}.

The European expert panel states in its executive summary that important progress has been made in demonstrating key aspects of plasma and dielectric accelerators, in particular in terms of energy and quality of the accelerated bunch from laser, electron and proton driven plasma accelerators. Also, rapid progress in underlying technologies, e.g., lasers, feedback systems, nano-control, manufacturing, etc., has been achieved. Given the state of development and work to be done the European expert panel and roadmap foresees that "a plasma-based collider can only become available for particle physics experiments beyond 2050. The feasibility of a collider based on plasma accelerator schemes remains to be proven." Consistent with the goals identified in this report, key challenges to reach the high energy frontier were identified and include the following aspects:
\begin{itemize}
    \item Demonstrate on paper and in experiment the acceleration of high electron bunch charge ($>$800 pC with good beam quality) in plasma, sufficient to reach the goal of specific luminosity.
    \item Develop on paper a scheme for multi-stage positron bunch acceleration in plasma, that can provide a sufficient bunch charge and emittance. Confirm in experiments.
    \item Develop emittance preservation at the nanometer scale, including measurement and correction procedures.
    \item Develop sufficient wall plug to beam power efficiency.
    \item Demonstrate staging designs of multiple structures with high energy gain and all optical elements, including tolerances and detailed length and cost scaling.
    \item Investigate in detail high repetition rate and associated power-handling and efficiency issues.
\end{itemize}

Given the strong potential gain with compact accelerators and the remaining challenges to be addressed, the panel recommended a strong work program with additional funding for HEP-oriented R\&D in novel accelerators. It was found that most progress has been achieved in plasma accelerators. Dielectric laser accelerators are in an earlier stage of development but are of potential interest on the longer horizon and are hence a part of the European future roadmap. Required additional resources for a European minimal plan amount to 147 FTE-years and 3.15 MCHF of investment. The aspirational plan would require additional resources for 147\,FTE-years and 35.5\,MCHF investment, beyond the minimal plan. 

The panel in Europe proposed a three-pillar R\&D roadmap for novel accelerators. A feasibility and pre-CDR study forms the first pillar and will investigate the potential and performance reach of plasma and laser accelerators for particle physics. In addition a realistic cost-size-benefit analysis is included and will be performed in a comparative approach for different technologies. A second pillar relies on technical demonstrations to prove suitability of plasma and laser accelerators for particle physics. A third pillar connects the work on novel accelerators to other science fields and to other applications. 

The European strategy for novel accelerators has explicitly reached out to collaborators in the US and Asia. It is clear that the work in Europe should be closely inter-linked and coordinated (or even integrated) with activities in the US or Asia. The European EuPRAXIA project includes US and Asian collaborators. The proposed "Feasibility and pre-CDR Study" is closely related to the integrated parameter sets and Integrated Design Study proposed in the Snowmass process. A common study or at least very close collaboration would leverage important synergy and should be envisaged. US accelerator R\&D facilities like BELLA and FACET-II offer unique and world-leading experimental capabilities. The European plan foresees that those capabilities are fully exploited and installations are not unnecessarily replicated in Europe. Such a collaborative, open innovation approach will follow the long tradition of particle physics. It will ensure most efficient use of resources and fastest progress. Adequate coordination, collaboration and funding schemes at the global level do, however, not exist for novel accelerators and remain to be designed and set up. Such structures could possibly be formed inside ICFA and adopting some ideas from the "Advanced LinEar collider study GROup" ALEGRO\cite{alegro.arxiv.2020}.

\newpage
\section{Near term applications as stepping stones towards high energy physics deployment}

Particle accelerators have been developed from the beginning of their century-long history to serve the needs of high energy physics community, pushing the energy and intensity frontiers in the search of better understanding of the subatomic world. At the same time, the capabilities of these machines to focus the electromagnetic energy into very high quality beams and the characteristics of the radiation emitted by these beams has been recognized early on and a myriad of uses of accelerators have been developed extending from pure scientific (material studies through electron or x-ray techniques such as microscopy, diffraction or spectroscopy, gamma rays for nuclear science) to more industrial (material processing, sterilization, nondestructive evaluation, security and nuclear nonproliferation, defense and space applications) to health sciences (radiation and therapy) applications. 

The field of high gradient advanced accelerators followed a similar trajectory: born in recognition of fundamental limits of conventional RF-based accelerators, in the quest for higher particle energies. In recent years,  an abundance of applications of compact particle accelerators and their associated radiation sources (electron, protons, X-rays, gammas, etc.) has emerged. Not only do advanced concepts shrink  the accelerator size and cost, but they enable entirely new frequencies and time-scales and corresponding radiation pulses. Controlling and diagnosing matter on ultra-small and ultra-short scales has been highlighted as a key driver of innovation and discovery \cite{BRNmanufacturing2020}. 

Building and developing advanced accelerators for near term application is critical to advance the technology as described in \cite{Emma22, Boucher22, emma2022snowmass}. Deferring real world use of this technology to very high energy collider-like machines will not facilitate  testing and solution of the myriad  technical challenges that separate an accelerator physics experiment from a working machine. Improving  output beam quality,  drive beam control, and energy transfer efficiency will only take place if actual users of plasma and laser based accelerators put the time in to perfect these machines. A  telling example is XFELs where a vibrant user community pushes the accelerator technology to frontiers that were  not foreseeable two decades ago. US funding agencies including NSF should recognize that fostering applications of advanced accelerators to other branches other than high energy physics would strongly enhance the possibilities for development of the technology. While advanced accelerator have significant ground to cover with respect to conventional RF-based accelerators, real-world deployment at energies lower than those needed for HEP colliders and application is an important step before the wider HEP community can entrust this technology with its future. This will also provide important societal benefits, e.g., in bringing diagnostic capabilities now only accessible in large facilities to clinical and industrial settings.   

Note that at the international level, strong competition has emerged for accelerator development for near-term applications. For example, two plasma-accelerator milestone demonstrations of free-electron lasing \cite{wang2021, pompili2022}, known to critically depend on excellent electron beam phase space characteristics, highlight recent advances.
In addition, the 2022 European particle physics report has specifically called out near-term applications (see ``Integration and Outreach" in Fig. 4.1 of \cite{EuroRoadmap2022}) as critical in achieving the long-term roadmap goals, which has led to concrete investments. The Eupraxia project \cite{Assman2020} is an example of a community-driven merging of accelerator concept and application needs. It cannot be ignored that while the field of plasma accelerator was first born and developed in the US, the two aforementioned plasma-based FEL demonstrations occurred in Asian and European facilities. A recent report from the BESAC Subcommittee on International Benchmarking (of US competitiveness) found similar trends and indicates possible mitigation strategies to increase competitiveness and retain scientific talent in other branches of basic energy sciences \cite{besac_report}.

The need for application-oriented facilities can be divided into two categories. For those where significant R\&D is still needed (for example, plasma-based and other advanced X-ray FEL concepts), it would be critical to obtain financial support for novel capabilities and beamlines at existing and new facilities. For those applications where proof-of-principle concepts have matured already (for example, betatron X-ray sources and laser-driven particle beams), a dedicated beamline or facility is needed where these sources can be permanently exploited, allowing both for efficient user-access as well as dedicated photon or particle source optimization. Many of the successes to date have been realized in short (few weeks long) campaigns, followed by beamline reconfiguration for other experiments. This has hampered source optimization, stability improvements, and integration of advanced transport and sample technology.

Table~\ref{tab:my_label} summarizes a selection of key near-term applications that are, or will be, enabled by advanced  accelerator concepts in the next 5-10 years. These span broadly across light source, medical and fundamental science applications and are described in detail in \cite{near-term-Snowmass21}. 

Following the previous 2013 Snowmass meeting,  the relevance of near-term applications of accelerators was  discussed in the context of communication and awareness among the general public (Section 10.4.2 in Ref. \cite{snowmass2013} states ``{\it Particle physics research has had a significant impact on other areas of science. Examples include applications developed for health and medicine...}"). This sentiment was echoed in the 2014 P5 report \cite{P5-2014} through ``{\it In return, developments within the particle physics community have enabled basic scientific research and applications in numerous other areas. This broad, connected scientific enterprise provides tremendous benefits to society as a whole.}" Stronger cross-agency emphasis on near-term applications of advanced accelerator concepts is critical  to society in general and also to the robustness of accelerator technology.  Such cross-agency engagement and funding for applications will support both broad development as well as the long-range particle physics mission. A successful near-term application environment will naturally guide particle accelerator technology to maturity. Applications demand robust day-to-day performance with minimal start-up time, control over parameters such as particle flux and energy, active stabilization concepts to keep the "up-time" large throughout the day, pump-probe synchronization, and reduction in operational expenses such as energy usage and component replacements, among others.

This was recognized in the 2016 Advanced Accelerator Development strategic report, \cite{roadmap.doe.2016} which put near-term applications on the high-energy physics roadmap, and should be included in the 2021 Snowmass and 2022 P5 reports. Furthermore, we emphasize that the near term applications discussed here enable opportunities for cross-cutting development in other Accelerator Frontier priorities (e.g., novel RF technology discussed in these proceedings \cite{nanni:ccc}) as well as in the Community Engagement Frontier (e.g. industrial applications based on FLASH radiation therapy in these proceedings \cite{Boucher22}).


\begin{table}
    \centering
    \scalebox{0.75}{
 \begin{tabular}{|p{0.3\linewidth}|>{\centering\arraybackslash}p{0.3\linewidth}|>{\centering\arraybackslash}p{0.3\linewidth}|>{\centering\arraybackslash}p{0.3\linewidth}|}
    \hline
         \textbf{Source} & \textbf{Example application} & \textbf{Status} & \textbf{Readiness in 5-10 years}\\ \hline
        Plasma-based FEL \cite{Emma2021} & Single-shot high-res imaging, non-linear excitation & Experimental feasibility demonstrated, two high-impact papers in 2021 & Realistic at higher flux and photon energy in $<$ 5 years  \\ \hline
        Corrugated-structure FEL \cite{Zholents2020} & Medical imaging & Conceptual Development  &  Technology to be explored\\ \hline
        Cryo-cooled Copper FEL  \cite{Rosenzweig2020} & Ultrafast Imaging, Attosecond Science & Conceptual Design & Technology to be explored\\ \hline
        Betatron X-rays  \cite{Corde_RMP_2013,Albert_PPCF_2016} &  Single-shot phase-contrast imaging of micro-structures    & Extensive demonstrations     &  Ready now  \\ \hline
        Compton-scattered X-rays \cite{Corde_RMP_2013,Albert_PPCF_2016} & Compact dose-reduced medical imaging, security, HED/IFE dynamics &   Proof-of-principle sources demonstrating applications including advanced imaging & Tunable and mono-energetic in $<$ 5 years  \\ \hline
        Advanced gamma ray sources \cite{Glinec_PRL_2005,Benismail_NIMA_2011,Gadjev2019,Sudar2020} & Security, efficient imaging at reduced dose & Experimental demonstrations (plasma based).
        Conceptual Development (non-plasma based) & Plasma-based ready now. Non-plasma based technology to be explored\\ \hline
        VHEE \cite{Svendsen2021,Labate2020,Kokurewicz2019} & Low dose radiotherapy & Well established, needs stability emphasis &  Ready now at compact low rep rate sources \\ \hline
        Laser-solid ions  \cite{Linz2016}&  Medical imaging, FLASH therapy, HED diagnostics, material science & Extensive demonstrations in TNSA regime & Ready now, $>$100 MeV protons in $<$5 years\\ \hline
        High-energy particle beams \cite{gessner2020beamdump} & Beam-dump explorations, astrophysical plasmas  & Initial experiments planned  & Results from initial experiments expected in $\sim$5 years\\ \hline
    \end{tabular}}
    \caption{High-level summary of near term applications of advanced accelerators described in the manuscript. References, example applications, status and readiness in 5-10 years are included for each source.}
    \label{tab:my_label}
\end{table}

\clearpage
\section{Facilities}

Demonstrating the viability of emerging accelerator science ultimately relies on experimental validation. A portfolio of beam test facilities at US national laboratories and universities, as well as international facilities in Europe and Asia, are used to perform research critical to advancing accelerator science and technology (S\&T). These facilities have enabled the pioneering accelerator research necessary to develop the next generation of accelerators. The current portfolio of U.S. beam test facilities and an overview of their research missions, recent achievements, and the upgrades required to keep the US competitive in light of the large investments in accelerator research around the world is given in \cite{BeamTestFacilities_Snowmass2021}.
The mission of the US beam test facilities is carried out by a broad community of accelerator scientists, drawn from universities, industry, and national laboratories and has three core aims: 
\begin{enumerate}
\item Provide the experimental testbeds where (1) exploratory accelerator research can be conducted (2) emerging accelerator concepts can be validated and (3) promising accelerator technology can be programmatically developed and undergo integration testing. 
\item Develop accelerator S\&T needed to enable the next generation of accelerator based HEP facilities for the energy- and intensity frontiers as well as other SC facilities such as light sources. 
\item Educate and train future scientists and engineers. 
\end{enumerate}
US beam test facilities deploy state of the art capabilities to enable these aims, including: O(10) GeV energy electron beams, O(1) Petawatt drive lasers and multiple beams, O(1) Gigawatt RF power sources, high-quality charged particle sources (e.g., low emittance electron beams), advanced beam manipulation systems (e.g. nonlinear integrable optics, optical stochastic cooling, emittance exchange) and capabilities to develop AI/ML for accelerator science.

US beam test facilities have facilitated rapid progress since the previous Snowmass Process. In 2015, the P5 strategic plan \cite{P5-2014} made multiple recommendations for the HEP General Accelerator R\&D (GARD) program. Roadmaps were successfully formulated by the GARD community in response to P5 to clearly articulate the longer-term research needed to develop the next generation of energy-frontier and intensity-frontier machines. For example, the 2016 Advanced Acceleration Concepts (AAC) Roadmap identified a series of milestones, many of which were demonstrated at the US beam test facilities. The beam test facilities were instrumental in addressing many of these as detailed in Section 3. A comprehensive update on all of the GARD subpanel recommendations was presented at the 2019 November HEPAP meeting \cite{hepap-meeting-2019}. Highlights of progress made on the subpanel recommendations illustrate the importance of the GARD beam test facilities. Examples of milestones accomplished include progress in both the intensity and energy frontiers. Energy-Frontier linear collider accelerator science delivered multi-GeV/m accelerating gradients and multi-GeV energy gain in plasma accelerators using beam-drivers at FACET \cite{Litos2014,Litos2015PPCF} and laser-drivers at BELLA \cite{gonsalves2019petawatt}, multi-GeV/m positron acceleration at FACET \cite{Corde2015,Doche2017}, staging of two plasma accelerator modules at BELLA \cite{SteinkeStaging}, demonstration of controlled injection in a plasma wakefield accelerator \cite{Deng2019}, and record setting transformer ratios at AWA in both structures \cite{gao2018prl} and plasmas \cite{rousel2020}. Related work includes ATF experiments on shock wave monoenergetic ion acceleration \cite{palmer-2011}, nonlinear effects in inverse Compton scattering (red shift, higher harmonics) \cite{babzien-2006}, high-gain high-harmonic generation FEL \cite{yu-2000}, fundamental research of electron acceleration in dielectric in structures (e.g., using 3D woodpile structures \cite{hoang-2018} and using metamaterial structures \cite{picard22}), fundamental research on nonlinear Thomson scattering \cite{fruhling-2021}, and novel methods for controlled optical injection of electrons into wakefields \cite{golovin2018electron}.

Upgrades to US beam test facilities will require significant investment by HEP to ensure the US GARD program remains internationally competitive. In turn, upgraded beam test facilities are needed to enable DOE-HEP to build a cutting-edge program in the future energy and intensity frontiers. For example, the European Strategy for Particle Physics has committed considerable funds to accelerator R\&D in Europe. All US GARD beam test facilities have near-term upgrades underway. Proposals for mid-term upgrades have been developed to continue progress on the roadmaps and support from DOE-HEP is needed to realize these plans. These include: kBELLA at LBNL for high rep-rate, precision laser plasma acceleration and subsequent applications; restoring the positron capability at SLAC FACET-II to demonstrate collider compatible techniques for accelerating positrons in plasma; CO2 laser power upgrades at BNL ATF for acceleration of electrons and ions with near-IR wavelengths; bunker expansion and drive beam energy upgrade at ANL AWA for GeV/m structure wakefield acceleration. 

\newpage
\section{Other advanced concepts for high gradient acceleration}


In addition to the three main technologies organized into the DOE Advanced Accelerator Development Strategy and roadmaps~\cite{roadmap.doe.2016} as the core of the field (beam and laser driven plasmas, and beam driven structures), new concepts continue to emerge and offer potential to improve performance of future accelerators. In this section we discuss some of these novel schemes including laser-driven dielectric structures \cite{england2022laser}, plasmonics \cite{Sahai22} and nanostructured/channeling accelerators \cite{ariniello2022channeling}. This is based on the principle, as recognized in the European roadmap more recently and in the series of Alegro workshops few years back, that the field of advanced accelerators could symmetrically be organized by including  possibilities where either laser or charged particle beams can be the drivers and the wave can be supported either by plasma or by a near field structure (dielectric or plasmonic). 
 

Laser-driven structures have been left off the advanced accelerator strategy roadmap, but significant progress has been made in the last 10 years in developing laser-driven microstructure accelerators based on dielectric laser acceleration (DLA), plasmonically-enhanced metasurface laser accelerators (MLA), and various photonic crystal configurations, which we here collectively refer to as laser-driven structure-based accelerators (LSA). This is mostly due to generous funding from the Moore foundation (ACHIP program) which has enabled reaching key milestones such as GV/m scale gradients, generation of attosecond electron bunches and on-chip miniaturization of various accelerator components. 

As an advanced accelerator concept, LSA offers some unique advantages. The acceleration mechanism is inherently linear and occurs in a vacuum region in a static structure. In addition to the stability benefits this affords, the acceleration effect is inherently dependent on the phase of the laser field, which makes it possible to dynamically fine-tune accelerator performance by manipulation of the incident laser phase profile. The acceleration mechanism also works equally well for positrons and electrons. Due to its unique low-charge high repetition rate bunch format, this approach may provide a possible source technology for proposed fixed-target light dark matter searches using single few-GeV electrons. In addition, the projected beamsstrahlung energy loss for a multi-TeV collider scenario is in the few percent range, as opposed to tens of percents for conventional RF accelerators. Furthermore, the primary supporting technologies (solid state lasers and nanofabrication) are mature and already at or near the capabilities required for a full-scale accelerator based on this approach. The LSA approach could significantly lower the cost per GeV by leveraging commercial developments of the integrated circuit and telecommunications industries. This is similar to the way that accelerators were developed in the 1950's and 1960's where they tapped into technology developed during WWII, e.g., microwave communications/radar/weapons development.

Looking towards the future, while synergistic applications (i.e., those needing ultra-compact accelerators) provide an opportunity for cost-sharing the technology development, overall the LSA technology has had significantly lower funding levels than other advanced accelerator technologies and essentially no funding dedicated to closing technology gaps unique for multi-TeV linear colliders. This is partly due to the perception that there is not a credible technology path for a 30 MW linear collider beam due to the small, micron-sized apertures, even when dividing the main beam power into multiple, parallel accelerating channels. This needs to be addressed as early in the LSA technology roadmap as possible. In addition to high quality acceleration with large energy gain, key steps needed for a LSA linear collider development include the following near-term technology demonstrations: (1) low-cost high-efficiency dielectric structures with sufficient thermal conductivity and controllable wakefield effects; (2) focusing schemes with sufficiently high gradient that minimize beam interception but do not dilute the beam quality; and (3) low-cost, high-efficiency, and high-power drive lasers that have sub-cycle phase and timing control. Successful demonstration of these capabilities will show that the LSA technology is a credible alternative multi-TeV linear collider technology and provide motivation for a future multi-stage linear collider prototype.

Far-field laser based acceleration (IFEL) scales badly with energy, but this technology can be important for low energy and to boost efficiency conversion between electromagnetic waves and relativistic beams in optical energy recirculation schemes. 

In addition there are emergent ideas to engineer the medium and take advantage of internal structure to further enhance the acceleration process. For example, plasmonic modes offers the potential to achieve petavolts per meter fields, that could transform the current paradigm in collider development in addition to non-collider searches in fundamental physics. Petavolts per meter plasmonics relies on collective oscillations of the free-electron Fermi gas inherent in the conduction band of materials that have a suitable combination of constituent atoms and ionic lattice structure.  The conduction band free electron density $n_0$ at equilibrium can be as high as 10$^{24}$cm$^{-3}$ and electromagnetic fields of plasmonic modes are of the order of 10$\sqrt{n_0(10^{22}\text{cm}^{-3})}$~TV/m. Engineered materials not only allow highly tunable material properties but quite critically make it possible to overcome disruptive instabilities that dominate the interactions in bulk media. Due to rapid shielding by the free electron Fermi gas, dielectric effects are strongly suppressed. Because of the ionic lattice, the corresponding electronic energy bands and the free electron gas are governed by quantum mechanical effects, comparisons with plasmas are merely notional. Based on this framework, it is critical to address various challenges that underlie petavolts per meter plasmonics including stable excitation of plasmonic modes while accounting for their effects on the ionic lattice and the electronic energy band structure over femtosecond timescales. Plasmonic colliders may shape the future by bringing high energy physics at the tens of TeV to multi-PeV center-of-mass-energies within reach. 

Note that plasmonics is significantly different from solid plasmas. Unlike the Fermi electron gas in conductive solids, solid plasmas do not exist at equilibrium and are created by ablation of solids such as by heating solids with a high-intensity optical laser. There is no ionic lattice and as a result in solid plasmas, individual ions are uncorrelated. Consequently, in plasmas there is no energy band structure and the Fermi electron gas is irreversibly disrupted. Due to the highly randomized nature of ions in plasmas, the properties of plasma oscillations significantly diverge from that of plasmonic modes especially at higher densities. Solid plasmas are utilized for ion acceleration in addition to being proposed for schemes such as channeling acceleration of positively charged particles.

A different concept is related to accelerator schemes based on particle channeling. Charged particles that are injected along crystallographic direction of a crystal can undergo a series of glancing collisions with the ionic lattice. If these collisions with individual ions are capable of trapping the charged particles within the inter-planar space, this phenomena is referred to as crystal channeling. Because the basis of energy exchange process is individual collisions of charged particles with positively charged ionic sites, it favors positively charged particles \cite{sahai:disambiguation}. 

Channeling acceleration in solid-state plasma of crystals or nanostructures, e.g., carbon nanotubes (CNTs) or alumna honeycomb holes, also has the promise of ultrahigh accelerating gradients O(1-10 TeV/m), continuous focusing of channeling particles without need of staging, and ultimately small equilibrium beam emittances naturally obtained while accelerating. Several methods of wakefield excitation are being considered including ultra-short or micro-bunched electron beams and laser pulses, ion clusters, etc. While beams of muons are the most suitable for potential future high energy physics colliders based on the crystal PWA. Support of  experimental efforts augmented with corresponding PIC plasma simulation would be required to expand the ongoing R\&D program and achieve a milestone demonstration by the end of this decade of 1 GeV acceleration over 1 mm of solid crystal/CNT structure.

\pagebreak
\section{Conclusion and R\&D priorities} 

A vigorous continued program of R\&D on advanced accelerators is needed, continuing strong recent progress in key accelerator elements to enable future colliders (as well as other applications). Significant results have been obtained since the last P5 report, including the production of high quality electron bunches at 8~GeV from a single stage,  new positron acceleration demonstrations and concepts for high quality, proof of principle staging of two LPA modules,  hollow channels, novel injection techniques for ultra-high beam brightness, investigation of processes that stabilize beam break up, and new technologies for high-average-power, high-efficiency lasers. Conceptual parameters have been developed for future colliders. Potential issues such as instabilities have been addressed and there appear to be mitigation techniques such that these do not  prevent a future collider. Advanced accelerator advantages such as energy recovery and short beams have strong potential to benefit such future machines. The community is ready to both continue R\&D and develop an integrated collider parameter set to further guide the R\&D program and lead towards a collider integrated design study. While the US has significant leadership in the field and well positioned test facilities, $\$$B-class overseas investments mean that re-invigoration of the R\&D effort is important to maintain position in this fast developing field. 

Priority research should continue to address and update the Advanced Accelerator Development Strategy~\cite{roadmap.doe.2016}:
\begin{itemize}
\item Vigorous research on advanced accelerators including experimental, theoretical, and computational components, should be conducted as part of the General Accelerator R\&D program. This will advance the advanced accelerator R{\&}D roadmaps towards future high energy colliders, develop intermediate applications, and ensure international competitiveness. Priority directions include staging of multiple modules at multi-GeV, high efficiency stages, preservation of emittance for electrons and positrons, high fidelity phase space control, active feedback precision control, and shaped beams and deployment of advanced accelerator applications. 

\item A targeted R\&D program addressing high energy advanced accelerator-based colliders (e.g., to 15~TeV, with intermediate options) should develop integrated parameter sets in coordination with international efforts. This should detail components of the system and their interactions, such the injector, drivers, plasma source, beam cooling, and beam delivery system. This would set the stage for an integrated design study and a future conceptual design report, after the next Snowmass.  

\item{Research in near-term applications should be recognized as essential to, and providing leverage for, progress towards HEP colliders. The interplay and mutual interests in this area between Offices in DOE-SC including HEP, BES, FES and ARDAP as well as with NSF, NNSA, defense and other agencies should be strengthened to advance and leverage research activities aimed at real-world deployment of advanced accelerators.}

\item{Advanced accelerators should continue to play a key-role in workforce development and diversity in accelerator physics. University programs and graduate students greatly benefit from the scientific visibility of the advanced accelerator field. Access to user facilities for graduate students and early career researchers as well as formal and hands on training opportunities in advanced beam and accelerator physics should be continued and enhanced.}

\item Enhanced driver {R\&}D is needed to develop the efficient, high repetition rate, high average power laser and charged particle beam technology that will power advanced accelerators colliders and societal applications. 
 
\item Support of upgrades for Beam Test Facilities are needed to maintain progress on advanced accelerator Roadmaps. These include development of a high repetition rate facility, proposed as kBELLA, to support precision active feedback and high rate; independently controllable positrons to explore high quality acceleration, proposed at FACET-II; and implementation of a integrated SWFA demonstrator, proposed at AWA.

\item A study for a collider demonstration facility and physics experiments at an intermediate energy (ca. 20--80~GeV) should establish a plan that would demonstrate essential technology and provide a facility for physics experiments at intermediate energy.  

\item A DOE-HEP sponsored workshop in the near term should update and formalize the U.S. advanced accelerator strategy and roadmaps including updates to the 2016 AARDS Roadmaps, and to coordinate efforts.
\end{itemize}


This program is consistent with the recommendations of  previous reports to prioritize advanced accelerator research, including the previous P5 and HEPAP subcommittee reports, the European Strategy and Laboratory Directors Group reports, and several others.  Such a program will position the US and partner programs to realize the promise of a new generation of colliders with potential energy reach to or beyond the 10 TeV scale. In addition to the long term goal of a high energy collider, advanced accelerators can provide compact sources of particles and photons for a wide variety of near-term applications in science, medicine, and industry which will provide both societal benefit and leverage for high energy physics. At the same time, the highly published and innovative research will serve to attract young talent to the accelerator field and will be important in realizing a more diverse workforce.

\appendix
\newpage
\section{Summary bullet points}
\begin{itemize}
\item Advanced accelerators in beam and laser driven structures/plasmas offer future e+e- or $\gamma \gamma$ colliders to the 15 TeV range: combining high precision with energy reach. Gradients of 10 GeV/m and beyond (few TeV/km geometric). 

\item Compactness and high fields, enablers for e+e- colliders of up to 15 TeV with 10 GeV/m gradients. 

\item Ultra-bright beam generation, fast cooling. Short beams that increase Luminosity per unit Power (L/P), potential for practical energy recovery

\item Driver technologies (SRF linacs, high average power lasers) are developing to meet needs of future colliders

\item Strong progress since last Snowmass: 10 GeV class beams, positrons, and FEL-lasing demonstrating high beam quality.

\item Community ready to develop integrated collider parameter set, moving towards design studies: follows logically from 2016 Advanced Accelerator Development Strategy.

\item Competitive field with strong investment overseas including prioritization in European planning (European minimum ask for particle physics-oriented AAC effort of 147 FTE-years + 3.15 MCh).

\item International community recognizes critical value of near term applications as stepping stones for HEP and to leverage societal impact (EUPRAXIA, DESY, CERN).

\item Synergies with existing or near future machines (plasma lens, beam shaping, upgrade paths, etc.) should be leveraged. Efforts on electron, positron, laser sources, and beam delivery systems are particularly important in this regard.

\item Strong R\&D investment needed in next generation upgrades of the test facilities: kHz acceleration at KBELLA, positrons at FACET-II, CO2 laser upgrade at ATF, energy upgrade at AWA and proton injector at FAST/IOTA. 

\item Advanced accelerator community is at the forefront in scientific visibility and serves as a strong attraction for workforce development.

\item New concepts (nanoplasmonics, laser-driven structures) continue to emerge and continue to offer ever higher gradient in the future.

\item Facilities developed for advanced accelerators serve all novel accelerator research, even beyond the roadmap technologies of LWFA, PWFA, SWFA. They provide beams for testing novel concepts such as DLA and NPA, plasmonic accelerator and channeling-based schemes and train the new generation of accelerator scientists.  

\item Need for common project and thrust toward a collider Integrated Design Study and/or intermediate demonstrator. Potential to re-use facilities of nearer term linear colliders offering attractive physics path for the field.

\end{itemize}
 
\acknowledgments
We gratefully acknowledge the input of all of the members of the Accelerator Frontier 6 group of Snowmass, and of colleagues in other Accelerator, Energy, Community and other Frontiers, which went into this report. The report in particular draws on the many white papers submitted, as well as presentations at the Summer Study and preceding meetings, and we appreciate their authors. This work was supported by the Director, Office of Science, Office of High Energy Physics, of the U.S. Department of Energy under Contracts including No. DE-AC02-05CH11231 and DE-AC02-76SF00515,  and by the National Science Foundation. The U.S. Government retains and the publisher, by accepting the article for publication, acknowledges that the U.S. Government retains a non-exclusive, paid-up, irrevocable, world-wide license to publish or reproduce the published form of this manuscript, or allow others to do so, for U.S.
Government purposes.
 
\newpage

\bibliographystyle{unsrt}
\bibliography{bibliography}

\begin{thebibliography}{100}

\bibitem{adolphsen2022european}
C.~Adolphsen, D.~Angal-Kalinin, T.~Arndt, et~al.
\newblock European strategy for particle physics -- accelerator {R\&D} roadmap.
\newblock {\em arXiv:2201.07895}, 2022.

\bibitem{roadmap.doe.2016}
{DOE Advanced Accelerator Concepts Research Roadmap Workshop}.
\newblock {Advanced Accelerator Development Strategy Report}, 2 2016.

\bibitem{gonsalves2019petawatt}
AJ~Gonsalves, K~Nakamura, J~Daniels, C~Benedetti, C~Pieronek, TCH De~Raadt,
  S~Steinke, JH~Bin, SS~Bulanov, J~Van~Tilborg, et~al.
\newblock Petawatt laser guiding and electron beam acceleration to 8 gev in a
  laser-heated capillary discharge waveguide.
\newblock {\em Physical review letters}, 122(8):084801, 2019.

\bibitem{Litos2015PPCF}
Michael Litos, E~Adli, J~Allen, Weiming An, C~Clarke, Sébastien Corde,
  Christopher Clayton, Joel Frederico, S~Gessner, S~Green, Mark Hogan,
  Chandrashekhar Joshi, W.~Lu, K~Marsh, W.~Mori, Margaux Schmeltz,
  N.~Vafaei-Najafabadi, and Vitaly Yakimenko.
\newblock 9 gev energy gain in a beam-driven plasma wakefield accelerator.
\newblock {\em Plasma Physics and Controlled Fusion}, 58, 11 2015.

\bibitem{corde2015multi}
S{\'e}bastien Corde, E~Adli, JM~Allen, W~An, CI~Clarke, CE~Clayton,
  JP~Delahaye, J~Frederico, S~Gessner, SZ~Green, et~al.
\newblock Multi-gigaelectronvolt acceleration of positrons in a self-loaded
  plasma wakefield.
\newblock {\em Nature}, 524(7566):442--445, 2015.

\bibitem{Litos2014Nature}
M.~Litos, E.~Adli, W.~An, C.~I. Clarke, C.~E. Clayton, S.~Corde, J.~P.
  Delahaye, R.~J. England, A.~S. Fisher, J.~Frederico, S.~Gessner, S.~Z. Green,
  M.~J. Hogan, C.~Joshi, W.~Lu, K.~A. Marsh, W.~B. Mori, P.~Muggli,
  N.~Vafaei-Najafabadi, D.~Walz, G.~White, Z.~Wu, V~Yakimenko, and G.~Yocky.
\newblock High- efficiency acceleration of an electron beam in a plasma
  wakefield accelerator.
\newblock {\em Nature}, 515:7525, 2014.

\bibitem{darcy.nature.2022}
R~D’Arcy, J~Chappell, J~Beinortaite, S~Diederichs, G~Boyle, B~Foster,
  MJ~Garland, P~Gonzalez Caminal, CA~Lindstr{\o}m, G~Loisch, et~al.
\newblock Recovery time of a plasma-wakefield accelerator.
\newblock {\em Nature}, 603(7899):58--62, 2022.

\bibitem{steinke2016multistage}
S~Steinke, J~Van~Tilborg, C~Benedetti, CGR Geddes, CB~Schroeder, J~Daniels,
  KK~Swanson, AJ~Gonsalves, K~Nakamura, NH~Matlis, et~al.
\newblock Multistage coupling of independent laser-plasma accelerators.
\newblock {\em Nature}, 530(7589):190--193, 2016.

\bibitem{rousel2020}
R.~Roussel, G.~Andonian, W.~Lynn, K.~Sanwalka, R.~Robles, C.~Hansel, A.~Deng,
  G.~Lawler, J.~B. Rosenzweig, G.~Ha, J.~Seok, J.~G. Power, M.~Conde,
  E.~Wisniewski, D.~S. Doran, and C.~E. Whiteford.
\newblock Single shot characterization of high transformer ratio wakefields in
  nonlinear plasma acceleration.
\newblock {\em Phys. Rev. Lett.}, 124:044802, Jan 2020.

\bibitem{gao2018prl}
Q.~Gao, G.~Ha, C.~Jing, S.~P. Antipov, J.~G. Power, M.~Conde, W.~Gai, H.~Chen,
  J.~Shi, E.~E. Wisniewski, D.~S. Doran, W.~Liu, C.~E. Whiteford, A.~Zholents,
  P.~Piot, and S.~S. Baturin.
\newblock Observation of high transformer ratio of shaped bunch generated by an
  emittance-exchange beam line.
\newblock {\em Phys. Rev. Lett.}, 120:114801, Mar 2018.

\bibitem{Oshea16}
B.~D. O’Shea, G.~Andonian, S.~K. Barber, K.~L. Fitzmorris, S.~Hakimi,
  J.~Harrison, P.~D. Hoang, M.~J. Hogan, B.~Naranjo, O.~B. Williams,
  V.~Yakimenko, and J.~B. Rosenzweig.
\newblock {Observation of acceleration and deceleration in
  gigaelectron-volt-per-metre gradient dielectric wakefield accelerators}.
\newblock {\em Nature Communications}, 7:12763, 2016.

\bibitem{LindstromPRL2021}
C.~A. Lindstr\o{}m, J.~M. Garland, S.~Schr\"oder, L.~Boulton, G.~Boyle,
  J.~Chappell, R.~D'Arcy, P.~Gonzalez, A.~Knetsch, V.~Libov, G.~Loisch,
  A.~Martinez de~la Ossa, P.~Niknejadi, K.~P\~oder, L.~Schaper, B.~Schmidt,
  B.~Sheeran, S.~Wesch, J.~Wood, and J.~Osterhoff.
\newblock Energy-spread preservation and high efficiency in a plasma-wakefield
  accelerator.
\newblock {\em Phys. Rev. Lett.}, 126:014801, Jan 2021.

\bibitem{pompili2021energy}
R~Pompili, D~Alesini, MP~Anania, M~Behtouei, M~Bellaveglia, A~Biagioni,
  FG~Bisesto, M~Cesarini, E~Chiadroni, A~Cianchi, et~al.
\newblock Energy spread minimization in a beam-driven plasma wakefield
  accelerator.
\newblock {\em Nature Physics}, 17(4):499--503, 2021.

\bibitem{wang2021}
Wentao Wang, Ke~Feng, Lintong Ke, Changhai Yu, Yi~Xu, Rong Qi, Yu~Chen, Zhiyong
  Qin, Zhiyong Zhang, Zhijun Zhang, Ming Fang, Jiaqi Liu, Kang~nan Jiang, Hao
  Wang, Cheng Wang, Xiaojun Yang, Fenxiang Wu, Yuxin Leng, Jiansheng Liu, Ruxin
  Li, and Zhizhan Xu.
\newblock Free-electron lasing at 27 nanometres based on a laser wakefield
  accelerator.
\newblock {\em {Nature}}, {595}:516–520, July {2021}.

\bibitem{pompili2022}
R~Pompili, D~Alesini, MP~Anania, S~Arjmand, M~Behtouei, M~Bellaveglia,
  A~Biagioni, B~Buonomo, F~Cardelli, M~Carpanese, et~al.
\newblock Free-electron lasing with compact beam-driven plasma wakefield
  accelerator.
\newblock {\em Nature}, 605(7911):659--662, 2022.

\bibitem{SOLEIL:plasmaFEL}
Marie Labat, Jurjen Couperus~CabadaÄŸ, A.~Ghaith, Arie Irman, Anthony
  Berlioux, Philippe Berteaud, FrÃ©dÃ©ric Blache, Stefan Bock, FranÃ§ois
  Bouvet, Fabien Briquez, Yen-Yu Chang, SÃ©bastien Corde, Alexander Debus,
  Carlos Oliveira, Jean-Pierre Duval, Yannick Dietrich, Moussa Ajjouri,
  Christoph Eisenmann, Julien gautier, and Marie Couprie.
\newblock Seeded free-electron laser driven by a compact laser plasma
  accelerator, 05 2022.

\bibitem{ITF22}
Thomas Roser and et~al.
\newblock Report of the snowmass’21 collider implementation task force.

\bibitem{besac_report}
{\em BESAC report: Can the US compete in Basic Energy Sciences?}
\newblock 2021.
\newblock
  \url{https://science.osti.gov/-/media/bes/pdf/reports/2021/International_Benchmarking-Report.pdf}.

\bibitem{Bai22}
M.~Bai, W.~A. Barletta, D.~L. Bruhwiler, S.~Chattopadhyay, Y.~Hao, S.~Holder,
  J.~Holzbauer, Z.~Huang, K.~Harkay, Y.~K. Kim, X.~Lu, S.~M. Lund, N.~Neveu,
  P.~Ostroumov, J.~R. Patterson, P.~Piot, T.~Satogata, A.~Seryi, A.~K. Soha,
  and S.~Winchester.
\newblock Strategies in education, outreach, and inclusion to enhance the us
  workforce in accelerator science and engineering, 2022.

\bibitem{Arce-Lareta22}
Enrique Arce-Larreta, Ketevi Assamagan, Emanuela Barzi, Uta Bilow, Kenneth
  Cecire, Sijbrand de~Jong, Simone Donati, Steven Goldfarb, Joel Klammer,
  Azwinndini Muronga, and Maria Niland.
\newblock The necessity of international particle physics opportunities for
  american education, 2022.

\bibitem{Barzi22}
Erin~V. Hansen, Erica Smith, Deborah Bard, Matthew Bellis, Jessica Esquivel,
  Tiffany~R. Lewis, Cameron Geddes, Cindy Joe, Alex~G. Kim, Asmita Patel, and
  Vitaly Pronskikh.
\newblock Climate of the field: Snowmass 2021, 2022.

\bibitem{Hansen22}
Emanuela Barzi, S.~James Gates, and Roxanne Springer.
\newblock In search of excellence and equity in physics, 2022.

\bibitem{Esarey09}
E.~Esarey, C.~B. Schroeder, and W.~P. Leemans.
\newblock Physics of laser-driven plasma-based electron accelerators.
\newblock {\em Rev. Mod. Phys.}, 81:1229--1285, July--September 2009.

\bibitem{Hooker13}
S.~M. Hooker.
\newblock Developments in laser-driven plasma accelerators.
\newblock {\em Nature Photonics}, 7:775--782, 2013.

\bibitem{benedetti.arxiv.2022a}
C.~Benedetti, S.~S. Bulanov, E.~Esarey, C.~G.~R. Geddes, A.~J. Gonsalves,
  A.~Huebl, R.~Lehe, K.~Nakamura, C.~B. Schroeder, D.~Terzani, J.~van Tilborg,
  M.~Turner, J.~L. Vay, T.~Zhou, F.~Albert, J.~Bromage, E.~M. Campbell, D.~H.
  Froula, J.~P. Palastro, J.~Zuegel, D.~Bruhwiler, N.~M. Cook, B.~Cros, M.~C.
  Downer, M.~Fuchs, B.~A. Shadwick, S.~J. Gessner, M.~J. Hogan, S.~M. Hooker,
  C.~Jing, K.~Krushelnick, A.~G.~R. Thomas, W.~P. Leemans, A.~R. Maier,
  J.~Osterhoff, K.~Poder, M.~Thevenet, W.~B. Mori, M.~Palmer, J.~G. Power, and
  N.~Vafaei-Najafabadi.
\newblock Linear collider based on laser-plasma accelerators, arxiv:
  2203.08366, 2022.

\bibitem{Gessner22}
Spencer Gessner and et~al.
\newblock Snowmass white paper on beam-driven plasma linear colliders, 2022.

\bibitem{Muggli22}
P.~Muggli and AWAKE Collaboration.
\newblock White paper: Awake, plasma wakefield acceleration of electron bunches
  for near and long term particle physics applications, 2022.

\bibitem{Jing22}
Chunguang Jing, John Power, Jiahang Shao, Gwanghui Ha, Philippe Piot, Xueying
  Lu, Alexander Zholents, Alexei Kanareykin, Sergey Kuzikov, James~B.
  Rosenzweig, Gerard Andonian, Evgenya~Ivanovna Simakov, Janardan Upadhyay,
  Chuanxiang Tang, Richard~J Temkin, Emilio~Alessandro Nanni, and John
  Lewellen.
\newblock Continuous and coordinated efforts of structure wakefield
  acceleration (swfa) development for an energy frontier machine, 2022.

\bibitem{XueyingLu22}
Xueying Lu, Jiahang Shao, John Power, Chunguang Jing, Gwanghui Ha, Philippe
  Piot, Alexander Zholents, Richard Temkin, Michael Shapiro, Julian Picard,
  Bagrat Grigoryan, Chuanxiang Tang, Yingchao Du, Jiaru Shi, Hao Zha, Dao
  Xiang, Emilio Nanni, Brendan O'Shea, Yuri Saveliev, Thomas Pacey, James
  Rosenzweig, Gerard Andonian, Evgenya Simakov, Francois Lemery, Alex Murokh,
  Sergey Kutsaev, Alexander Smirnov, Ronald Agustsson, Gongxiaohui Chen,
  Seunghwan Shin, and Ben Freemire.
\newblock Advanced rf structures for wakefield acceleration and high-gradient
  research, 2022.

\bibitem{Fuchs22}
M.~Fuchs, B.~A. Shadwick, N.~Vafaei-Najafabadi, A.~G.~R. Thomas, G.~Andonian,
  M.~Büscher, A.~Lehrach, O.~Apsimon, G.~Xia, D.~Filippetto, C.~B. Schroeder,
  and M.~C. Downer.
\newblock Snowmass whitepaper af6: Plasma-based particle sources, 2022.

\bibitem{Gessner22b}
Spencer Gessner and et~al.
\newblock Beam delivery and final focus systems for multi-tev advanced linear
  colliders, 2022.

\bibitem{Kiani22}
Leily Kiani, Tong Zhou, Seung-Whan Bahk, Jake Bromage, David Bruhwiler,
  E.~Michael Campbell, Zenghu Chang, Enam Chowdhury, Michael Downer, Qiang Du,
  Eric Esarey, Almantas Galvanauskas, Thomas Galvin, Constantin Hafner, Dieter
  Hoffmann, Chan Joshi, Manoj Kanskar, Wei Lu, Carmen Menoni, Michael Messerly,
  Sergey~B. Mirov, Mark Palmer, Igor Pogorelsky, Mikhail Polyanskiy, Erik
  Power, Brendan Reagan, Jorge Rocca, Joshua Rothenberg, Bruno~E. Schmidt,
  Emily Sistrunk, Thomas Spinka, Sergei Tochitsky, Navid Vafaei-Najafabadi,
  Jeroen van Tilborg, Russell Wilcox, Jonathan Zuegel, and Cameron Geddes.
\newblock High average power ultrafast laser technologies for driving future
  advanced accelerators, 2022.

\bibitem{Emma22}
Tim Barklow, Su~Dong, Claudio Emma, Joseph Duris, Zhirong Huang, Adham Naji,
  Emilio Nanni, James Rosenzweig, Anne Sakdinawat, Sami Tantawi, and Glen
  White.
\newblock Xcc: An x-ray fel-based $\gamma-\gamma$ collider higgs factory, 2022.

\bibitem{Boucher22}
Salime Boucher, Eric Esarey, Cameron Geddes, Carol Johnstone, Sergey Kutsaev,
  Billy~W. Loo, Francois Méot, Brahim Mustapha, Kei Nakamura, Emilio Nanni,
  Lieselotte Obst-Huebl, Stephen~E. Sampayan, Carl Schroeder, Reinhard Schulte,
  Ke~Sheng, Antoine Snijders, Emma Snively, Sami~G. Tantawi, and Jeroen van
  Tilborg.
\newblock Transformative technology for flash radiation therapy: A snowmass
  2021 white paper, 2022.

\bibitem{England22}
R.~J. England, D.~Filippetto, G.~Torrisi, A.~Bacci, G.~Della~Valle, D.~Mascali,
  G.~S. Mauro, G.~Sorbello, P.~Musumeci, J.~Scheuer, B.~Cowan, L~Schachter,
  Y-C. Huang, U.~Niedermayer, W.~D. Kimura, R.~Li, R.~Ishebeck, E.~I. Simakov,
  P.~Hommelhoff, and R.~L. Byer.
\newblock Laser-driven structure-based accelerators, 2022.

\bibitem{Sahai22}
Aakash~A. Sahai, Mark Golkowski, Stephen Gedney, Thomas Katsouleas, Gerard
  Andonian, Glen White, Joachim Stohr, Patric Muggli, Daniele Filipetto, Frank
  Zimmermann, Toshiki Tajima, Gerard Mourou, and Javier Resta-Lopez.
\newblock Petavolts per meter plasmonics: Snowmass21 white paper, 2022.

\bibitem{Ariniello22}
Robert Ariniello, Sebastien Corde, Xavier Davoine, Henrik Ekerfelt, Frederico
  Fiuza, Max Gilljohann, Laurent Gremillet, Yuliia Mankovska, Henryk Piekarz,
  Pablo San~Miguel Claveria, Vladimir Shiltsev, Peter Taborek, and Toshiki
  Tajima.
\newblock Channeling acceleration in crystals and nanostructures and studies of
  solid plasmas: New opportunities, 2022.

\bibitem{P5-2014}
{\em Report of the Particle Physics Project Prioritization Panel (P5): Building
  for Discovery Strategic Plan for U.S. Particle Physics in the Global
  Context}. U.S.\ Department of Energy, 2014.

\bibitem{HEPAP15}
{\em Accelerating Discovery: A Strategic Plan for Accelerator R\&D in the U.S.}
  U.S.\ Department of Energy, 2015.

\bibitem{ESPPU20}
{\em 2020 Update of the European Strategy for Particle Physics}, number
  CERN-ESU-015. CERN, 2020.

\bibitem{LDG22}
{\em European Strategy for Particle Physics - Accelerator R\&D Roadmap}, number
  arXiv:2201.07895. CERN, 2022.

\bibitem{BRN19}
{\em Basic Research Needs Workshop on Compact Accelerators for Security and
  Medicine}. US Department of Energy, Office of Science, US Department of
  Energy, Office of Science, 2019.

\bibitem{BrightestLightReprot18}
{\em The Future of Intense Ultrafast Lasers in the U.S., Brightest Light
  Initiative Workshop Report}. Optical Society of America, 2019.

\bibitem{DPP20}
{\em A Community Plan for Fusion Energy and Discovery Plasma Sciences}.
  American Physical Society, Division of Plasma Physics, 2020.

\bibitem{NAS-Decadal20}
{\em Plasma Science: Enabling Technology, Sustainability, Security, and
  Exploration}. National Academies Press, 2020.

\bibitem{Geddes04}
C.~G.~R. Geddes, Cs. Toth, J.~van Tilborg, E.~Esarey, C.~B. Schroeder,
  D.~Bruhwiler, C.~Nieter, J.~Cary, and W.~P. Leemans.
\newblock High-quality electron beams from a laser wakefield accelerator using
  plasma-channel guiding.
\newblock {\em Nature}, 431:538--541, 2004.

\bibitem{Faure04}
J.~Faure, Y.~Glinec, A.~Pkhov, S.~Kiselev, S.~Gordienko, E.~Lefebvre, J.-P.
  Rousseau, F.~Burgy, and V.~Malka.
\newblock A laser–plasma accelerator producing monoenergetic electron beams.
\newblock {\em Nature}, 431:541--544, 2004.

\bibitem{Mangles04}
S.~P.~D. Mangles, C.~D. Murphy, Z.~Najmudin, A.~G.~R. Thomas, J.~L. Collier,
  A.~E. Dangor, E.~J. Divall, P.~S. Foster, J.~G. Gallacher, C.~J. Hooker,
  D.~A. Jaroszynski, A.~J. Langley, W.~B. Mori, P.~A. Norreys, F.~S. Tsung,
  R.~Viskup, B.~R. Walton, and K.~Krushelnick.
\newblock Monoenergetic beams of relativistic electrons from intense
  laser–plasma interactions.
\newblock {\em Nature}, 431:535--538, 2004.

\bibitem{Gonsalves19}
A.~J. Gonsalves, K.~Nakamura, J.~Daniels, C.~Benedetti, C.~Pieronek, T.~C.~H.
  de~Raadt, S.~Steinke, J.~H. Bin, S.~S. Bulanov, J.~van Tilborg, C.~G.~R.
  Geddes, C.~B. Schroeder, Cs. T\'oth, E.~Esarey, K.~Swanson, L.~Fan-Chiang,
  G.~Bagdasarov, N.~Bobrova, V.~Gasilov, G.~Korn, P.~Sasorov, and W.~P.
  Leemans.
\newblock Petawatt laser guiding and electron beam acceleration to {8 GeV} in a
  laser-heated capillary discharge waveguide.
\newblock {\em Phys. Rev. Lett.}, 122:084801, Feb 2019.

\bibitem{Shalloo19}
R.~J. Shalloo, C.~Arran, A.~Picksley, A.~von Boetticher, L.~Corner,
  J.~Holloway, G.~Hine, J.~Jonnerby, H.~M. Milchberg, C.~Thornton, R.~Walczak,
  and S.~M. Hooker.
\newblock Low-density hydrodynamic optical-field-ionized plasma channels
  generated with an axicon lens.
\newblock {\em Phys. Rev. Accel. Beams}, 22:041302, Apr 2019.

\bibitem{Miao20}
B.~Miao, L.~Feder, J.~E. Shrock, A.~Goffin, and H.~M. Milchberg.
\newblock Optical guiding in meter-scale plasma waveguides.
\newblock {\em Phys. Rev. Lett.}, 125:074801, Aug 2020.

\bibitem{Leemans09}
Wim Leemans and Eric Esarey.
\newblock Laser-driven plasma-wave electron accelerators.
\newblock {\em Physics Today}, 62(3):44--49, 2009.

\bibitem{Schroeder10b}
C.~B. Schroeder, E.~Esarey, C.~G.~R. Geddes, C.~Benedetti, and W.~P. Leemans.
\newblock Physics considerations for laser-plasma linear colliders.
\newblock {\em Phys. Rev. ST Accel. Beams}, 13(10):101301, Oct 2010.

\bibitem{Litos2014}
M.~Litos et~al.
\newblock High-efficiency acceleration of an electron beam in a plasma
  wakefield accelerator.
\newblock {\em Nature}, 515(7525):92--95, November 2014.

\bibitem{kuzikov:ipac2021-wepab163}
S.V. Kuzikov, S.P. Antipov, P.V. Avrakhov, E.~Dosov, G.~Ha, C.-J. Jing, E.W.
  Knight, W.~Liu, X.~Lu, P.~Piot, J.G. Power, D.S. Scott, J.H. Shao, W.H. Tan,
  and E.E. Wisniewski.
\newblock {An X-Band Ultra-High Gradient Photoinjector}.
\newblock In {\em Proc. IPAC'21}, number~12 in International Particle
  Accelerator Conference, pages 2986--2989. JACoW Publishing, Geneva,
  Switzerland, 08 2021.
\newblock https://doi.org/10.18429/JACoW-IPAC2021-WEPAB163.

\bibitem{abp2021arxiv}
S.~Nagaitsev, Z.~Huang, J.~Power, J.~L. Vay, P.~Piot, L.~Spentzouris,
  J.~Rosenzweig, Y.~Cai, S.~Cousineau, M.~Conde, M.~Hogan, A.~Valishev,
  M.~Minty, T.~Zolkin, X.~Huang, V.~Shiltsev, J.~Seeman, J.~Byrd, Y.~Hao,
  B.~Dunham, B.~Carlsten, A.~Seryi, and R.~Patterson.
\newblock Accelerator and beam physics research goals and opportunities, 2021.

\bibitem{SteinkeStaging}
S.~Steinke et~al.
\newblock {Multistage coupling of independent laser-plasma accelerators}.
\newblock {\em Nature}, 530:190--193, 2016.

\bibitem{JING201872}
C.~Jing, S.~Antipov, M.~Conde, W.~Gai, G.~Ha, W.~Liu, N.~Neveu, J.G. Power,
  J.~Qiu, J.~Shi, D.~Wang, and E.~Wisniewski.
\newblock Electron acceleration through two successive electron beam driven
  wakefield acceleration stages.
\newblock {\em Nuclear Instruments and Methods in Physics Research Section A:
  Accelerators, Spectrometers, Detectors and Associated Equipment}, 898:72--76,
  2018.

\bibitem{Corde2015}
S.~Corde et~al.
\newblock Multi-gigaelectronvolt acceleration of positrons in a self-loaded
  plasma wakefield.
\newblock {\em Nature}, 524(7566):442--445, August 2015.

\bibitem{Gessner2016}
{Demonstration of a positron beam-driven hollow channel plasma wakefield
  accelerator}.
\newblock {\em Nature Communications}, 7:5--10, 2016.

\bibitem{Schroeder16}
C.B. Schroeder, C.~Benedetti, E.~Esarey, and W.P. Leemans.
\newblock Laser-plasma-based linear collider using hollow plasma channels.
\newblock {\em Nucl. Instrum. Methods Phys. Res. A}, 829:113--116, 2016.

\bibitem{Lindstrom18posi}
C.~A. Lindstr\o{}m, E.~Adli, J.~M. Allen, W.~An, C.~Beekman, C.~I. Clarke,
  C.~E. Clayton, S.~Corde, A.~Doche, J.~Frederico, S.~J. Gessner, S.~Z. Green,
  M.~J. Hogan, C.~Joshi, M.~Litos, W.~Lu, K.~A. Marsh, W.~B. Mori, B.~D.
  O'Shea, N.~Vafaei-Najafabadi, and V.~Yakimenko.
\newblock Measurement of transverse wakefields induced by a misaligned positron
  bunch in a hollow channel plasma accelerator.
\newblock {\em Phys. Rev. Lett.}, 120:124802, Mar 2018.

\bibitem{Diederichs2019}
S.~Diederichs et~al.
\newblock Positron transport and acceleration in beam-driven plasma wakefield
  accelerators using plasma columns.
\newblock {\em Physical Review Accelerators and Beams}, 22(8), August 2019.

\bibitem{Adli2018}
E.~Adli et~al.
\newblock Acceleration of electrons in the plasma wakefield of a proton bunch.
\newblock {\em Nature}, 561(7723):363--367, August 2018.

\bibitem{Doss2019}
C.{\hspace{0.167em}}E. Doss, E.~Adli, R.~Ariniello, J.~Cary, S.~Corde,
  B.~Hidding, M.{\hspace{0.167em}}J. Hogan, K.~Hunt-Stone, C.~Joshi,
  K.{\hspace{0.167em}}A. Marsh, J.{\hspace{0.167em}}B. Rosenzweig,
  N.~Vafaei-Najafabadi, V.~Yakimenko, and M.~Litos.
\newblock Laser-ionized, beam-driven, underdense, passive thin plasma lens.
\newblock {\em Physical Review Accelerators and Beams}, 22(11), November 2019.

\bibitem{chen:1989prd}
P.~Chen, S.~Rajagopalan, and J.~Rosenzweig.
\newblock Final focusing and enhanced disruption from an underdense plasma lens
  in a linear collider.
\newblock {\em Physical Review D}, 40(3):923--926, August 1989.

\bibitem{chen:1990prl}
P.~Chen, K.~Oide, A.~M. Sessler, and S.~S. Yu.
\newblock Plasma-based adiabatic focuser.
\newblock {\em Physical Review Letters}, 64(11):1231--1234, March 1990.

\bibitem{vanTilborg15}
J.~van Tilborg, S.~Steinke, C.~G.~R. Geddes, N.~H. Matlis, B.~S. Shaw, A.~J.
  Gonsalves, J.~V. Huijts, K.~Nakamura, J.~Daniels, C.~B. Schroeder, S.~S.
  Bulanov, N.~A. Bobrova, P.~V. Sasorov, and W.~P. Leemans.
\newblock Active plasma lensing for relativistic laser-plasma-accelerated
  electron beams.
\newblock {\em Phys. Rev. Lett.}, 115:184802, 2015.

\bibitem{Pompili18}
R.~Pompili, M.~P. Anania, M.~Bellaveglia, A.~Biagioni, S.~Bini, F.~Bisesto,
  E.~Brentegani, F.~Cardelli, G.~Castorina, E.~Chiadroni, A.~Cianchi, O.~Coiro,
  G.~Costa, M.~Croia, D.~Di Giovenale, M.~Ferrario, F.~Filippi, A.~Giribono,
  V.~Lollo, A.~Marocchino, M.~Marongiu, V.~Martinelli, A.~Mostacci,
  D.~Pellegrini, L.~Piersanti, G.~Di Pirro, S.~Romeo, A.~R. Rossi, J.~Scifo,
  V.~Shpakov, A.~Stella, C.~Vaccarezza, F.~Villa, and A.~Zigler.
\newblock Focusing of high-brightness electron beams with active-plasma lenses.
\newblock {\em Phys. Rev. Lett.}, 121:174801, 2018.

\bibitem{Lindstrom18}
C.~A. Lindstrom, E.~Adli, G.~Boyle, R.~Corsini, A.~E. Dyson, W.~Farabolini,
  S.~M. Hooker, M.~Meisel, J.~Osterhoff, J.-H. R{\"o}ckemann, L.~Schaper, and
  K.~N. Sjobak.
\newblock Emittance preservation in an aberration-free active plasma lens.
\newblock {\em Phys. Rev. Lett.}, 121:194801, 2018.

\bibitem{Lindstrm2018}
C.{\hspace{0.167em}}A. Lindstr{\o}m et~al.
\newblock Emittance preservation in an aberration-free active plasma lens.
\newblock {\em Physical Review Letters}, 121(19), November 2018.

\bibitem{Bulanov98}
S.~Bulanov, N.~Naumova, F.~Pegoraro, and J.~Sakai.
\newblock Particle injection into the wave acceleration phase due to nonlinear
  wake wave breaking.
\newblock {\em Phys. Rev. E}, 58:R5257--R5260, Nov 1998.

\bibitem{Geddes08}
C.~G.~R. Geddes, K.~Nakamura, G.~R. Plateau, Cs. Toth, E.~Cormier-Michel,
  E.~Esarey, C.~B. Schroeder, J.~R. Cary, and W.~P. Leemans.
\newblock Plasma-density-gradient injection of low absolute-momentum-spread
  electron bunches.
\newblock {\em Phys. Rev. Lett.}, 100:215004, May 2008.

\bibitem{Gonsalves11}
A.~J. Gonsalves, K.~Nakamura, C.~Lin, D.~Panasenko, S.~Shiraishi, T.~Sokollik,
  C.~Benedetti, C.~B. Schroeder, C.~G.~R. Geddes, J.~van Tilborg, J.~Osterhoff,
  E.~Esarey, C.~Toth, and W.~P. Leemans.
\newblock Tunable laser plasma accelerator based on longitudinal density
  tailoring.
\newblock {\em Nature Phys.}, 7:862--866, 2011.

\bibitem{Goetzfried20}
J.~G\"otzfried, A.~D\"opp, M.~F. Gilljohann, F.~M. Foerster, H.~Ding,
  S.~Schindler, G.~Schilling, A.~Buck, L.~Veisz, and S.~Karsch.
\newblock Physics of high-charge electron beams in laser-plasma wakefields.
\newblock {\em Phys. Rev. X}, 10:041015, Oct 2020.

\bibitem{Deng2019}
A.~Deng et~al.
\newblock Generation and acceleration of electron bunches from a plasma
  photocathode.
\newblock {\em Nature Physics}, 15(11):1156--1160, August 2019.

\bibitem{KnetschTorchPRAB2020}
D.~Ullmann, P.~Scherkl, A.~Knetsch, T.~Heinemann, A.~Sutherland, A.~F. Habib,
  O.~S. Karger, A.~Beaton, G.~G. Manahan, A.~Deng, G.~Andonian, M.~D. Litos,
  B.~D. O'Shea, J.~R. Cary, M.~J. Hogan, V.~Yakimenko, J.~B. Rosenzweig, and
  B.~Hidding.
\newblock All-optical density downramp injection in electron-driven plasma
  wakefield accelerators.
\newblock {\em Phys. Rev. Research}, 3:043163, Dec 2021.

\bibitem{XuPRABdownramp2017}
X.~L. Xu, F.~Li, W.~An, T.~N. Dalichaouch, P.~Yu, W.~Lu, C.~Joshi, and W.~B.
  Mori.
\newblock High quality electron bunch generation using a longitudinal
  density-tailored plasma-based accelerator in the three-dimensional blowout
  regime.
\newblock {\em Phys. Rev. Accel. Beams}, 20:111303, Nov 2017.

\bibitem{UllmannPhysRevResearch2021}
D.~Ullmann, P.~Scherkl, A.~Knetsch, T.~Heinemann, A.~Sutherland, A.~F. Habib,
  O.~S. Karger, A.~Beaton, G.~G. Manahan, A.~Deng, G.~Andonian, M.~D. Litos,
  B.~D. O'Shea, J.~R. Cary, M.~J. Hogan, V.~Yakimenko, J.~B. Rosenzweig, and
  B.~Hidding.
\newblock All-optical density downramp injection in electron-driven plasma
  wakefield accelerators.
\newblock {\em Phys. Rev. Research}, 3:043163, Dec 2021.

\bibitem{Wittig2015Downramp}
G.~Wittig, O.~Karger, A.~Knetsch, Y.~Xi, A.~Deng, J.~B. Rosenzweig, D.~L.
  Bruhwiler, J.~Smith, G.~G. Manahan, Z.~M. Sheng, et~al.
\newblock Optical plasma torch electron bunch generation in plasma wakefield
  accelerators.
\newblock {\em Phys. Rev. ST Accel. Beams}, 18(8):081304, 2015.

\bibitem{CouperusDownrampPRR2021}
J.~P. Couperus Cabada\ifmmode~\breve{g}\else \u{g}\fi{}, R.~Pausch,
  S.~Sch\"obel, M.~Bussmann, Y.-Y. Chang, S.~Corde, A.~Debus, H.~Ding,
  A.~D\"opp, F.~M. Foerster, M.~Gilljohann, F.~Haberstroh, T.~Heinemann,
  B.~Hidding, S.~Karsch, A.~Koehler, O.~Kononenko, A.~Knetsch, T.~Kurz,
  A.~Martinez de~la Ossa, A.~Nutter, G.~Raj, K.~Steiniger, U.~Schramm, P.~Ufer,
  and A.~Irman.
\newblock Gas-dynamic density downramp injection in a beam-driven plasma
  wakefield accelerator.
\newblock {\em Phys. Rev. Research}, 3:L042005, Oct 2021.

\bibitem{Esarey97}
E.~Esarey, R.~F. Hubbard, W.~P. Leemans, A.~Ting, and P.~Sprangle.
\newblock Electron injection into plasma wakefields by colliding laser pulses.
\newblock {\em Phys. Rev. Lett.}, 79:2682--2685, Oct 1997.

\bibitem{Faure07}
J.~Faure, C.~Rechatin, A.~Norlin, A.~Lifschitz, Y.~Glinec, and V.~Malka.
\newblock Controlled injection and acceleration of electrons in plasma
  wakefields by colliding laser pulses.
\newblock {\em Nature}, 444:737--739, 2007.

\bibitem{Rechatin09}
C.~Rechatin, J.~Faure, A.~Ben-Ismail, J.~Lim, R.~Fitour, A.~Specka, H.~Videau,
  A.~Tafzi, F.~Burgy, and V.~Malka.
\newblock Controlling the phase-space volume of injected electrons in a
  laser-plasma accelerator.
\newblock {\em Phys. Rev. Lett.}, 102:164801, Apr 2009.

\bibitem{Geddes16}
{\em High energy, low energy spread electron bunches produced via colliding
  pulse injection}, volume 1777, 2016.

\bibitem{Kirchen21}
Manuel Kirchen, S\"oren Jalas, Philipp Messner, Paul Winkler, Timo Eichner,
  Lars H\"ubner, Thomas H\"ulsenbusch, Laurids Jeppe, Trupen Parikh, Matthias
  Schnepp, and Andreas~R. Maier.
\newblock Optimal beam loading in a laser-plasma accelerator.
\newblock {\em Phys. Rev. Lett.}, 126:174801, Apr 2021.

\bibitem{Ke2021}
L.~T. Ke, K.~Feng, W.~T. Wang, Z.~Y. Qin, C.~H. Yu, Y.~Wu, Y.~Chen, R.~Qi,
  Z.~J. Zhang, Y.~Xu, X.~J. Yang, Y.~X. Leng, J.~S. Liu, R.~X. Li, and Z.~Z.
  Xu.
\newblock Near-gev electron beams at a few per-mille level from a laser
  wakefield accelerator via density-tailored plasma.
\newblock {\em Phys. Rev. Lett.}, 126:214801, May 2021.

\bibitem{Plateau12}
G.~R. Plateau, C.~G.~R. Geddes, D.~B. Thorn, M.~Chen, C.~Benedetti, E.~Esarey,
  A.~J. Gonsalves, N.~H. Matlis, K.~Nakamura, C.~B. Schroeder, S.~Shiraishi,
  T.~Sokollik, J.~van Tilborg, Cs. Toth, S.~Trotsenko, T.~S. Kim, M.~Battaglia,
  Th. St\"ohlker, and W.~P. Leemans.
\newblock Low-emittance electron bunches from a laser-plasma accelerator
  measured using single-shot x-ray spectroscopy.
\newblock {\em Phys. Rev. Lett.}, 109:064802, Aug 2012.

\bibitem{Weingartner12}
R.~Weingartner, S.~Raith, A.~Popp, S.~Chou, J.~Wenz, K.~Khrennikov,
  M.~Heigoldt, A.~R. Maier, N.~Kajumba, M.~Fuchs, B.~Zeitler, F.~Krausz,
  S.~Karsch, and F.~Gr\"uner.
\newblock Ultralow emittance electron beams from a laser-wakefield accelerator.
\newblock {\em Phys. Rev. ST Accel. Beams}, 15:111302, Nov 2012.

\bibitem{Lundh11}
O.~Lundh, J.~Lim, C.~Rechatin, L.~Ammoura, A.~Ben-Ismail, X.~Davoine,
  G.~Gallot, {J.-P.} Goddet, E.~Lefebvre, V.~Malka, and J.~Faure.
\newblock Few femtosecond, few kiloampere electron bunch produced by a
  laser--plasma accelerator.
\newblock {\em Nature Phys.}, 7:219---222, 2011.

\bibitem{Buck11}
Alexander Buck, Maria Nicolai, Karl Schmid, Chris M.~S. Sears, Alexander
  S\"{a}vert, Julia~M. Mikhailova, Ferenc Krausz, Malte~C. Kaluza, and Laszlo
  Veisz.
\newblock Real-time observation of laser-driven electron acceleration.
\newblock {\em Nature Phys.}, 7:543--548, March 2011.

\bibitem{cuperus17}
J.~P. Couperus, R.~Pausch, O.~K\"ohler, A. ad~Zarini, J.~M. Kr\"amer,
  M.~Garten, A.~Huebl, R.~Gebhardt, U.~Helbig, S.~Bock, K.~Zeil, A.~Debus,
  M.~Bussmann, U.~Schramm, and A.~Irman.
\newblock Demonstration of a beam loaded nanocoulomb-class laser wakefield
  accelerator.
\newblock {\em Nature Commun.}, 8:487, 2017.

\bibitem{Wu19}
Yitong Wu, Liangliang Ji, Xuesong Geng, Qin Yu, Nengwen Wang, Bo~Feng, Zhao
  Guo, Weiqing Wang, Chengyu Qin, Xue Yan, Lingang Zhang, Johannes Thomas, Anna
  Hützen, Markus Büscher, T~Peter Rakitzis, Alexander Pukhov, Baifei Shen,
  and Ruxin Li.
\newblock Polarized electron-beam acceleration driven by vortex laser pulses.
\newblock {\em New Journal of Physics}, 21(7):073052, jul 2019.

\bibitem{Wen19}
Meng Wen, Matteo Tamburini, and Christoph~H. Keitel.
\newblock Polarized laser-wakefield-accelerated kiloampere electron beams.
\newblock {\em Phys. Rev. Lett.}, 122:214801, May 2019.

\bibitem{Rykovanov15}
S.~G. Rykovanov, C.~B. Schroeder, E.~Esarey, C.~G.~R. Geddes, and W.~P.
  Leemans.
\newblock Plasma undulator based on laser excitation of wakefields in a plasma
  channel.
\newblock {\em Phys. Rev. Lett.}, 114:145003, Apr 2015.

\bibitem{Wang17}
J~W Wang, C~B Schroeder, R~Li, M~Zepf, and S~G Rykovanov.
\newblock Plasma channel undulator excited by high-order laser modes.
\newblock {\em Sci Rep}, 7(1):16884, Dec 2017.

\bibitem{Zhou:15}
Tong Zhou, John Ruppe, Cheng Zhu, I-Ning Hu, John Nees, and Almantas
  Galvanauskas.
\newblock {Coherent pulse stacking amplification using low-finesse
  Gires-Tournois interferometers}.
\newblock {\em Opt. Express}, 23(6):7442--7462, Mar 2015.

\bibitem{Zhou:18}
Tong Zhou, Qiang Du, Tyler Sano, Russell Wilcox, and Wim Leemans.
\newblock Two-dimensional combination of eight ultrashort pulsed beams using a
  diffractive optic pair.
\newblock {\em Opt. Lett.}, 43(14):3269--3272, Jul 2018.

\bibitem{Chang:13}
W.-Z. Chang, Tong Zhou, Leo~A. Siiman, and Almantas Galvanauskas.
\newblock Femtosecond pulse spectral synthesis in coherently-spectrally
  combined multi-channel fiber chirped pulse amplifiers.
\newblock {\em Opt. Express}, 21(3):3897--3910, Feb 2013.

\bibitem{Galvin19}
T.~C. Galvin, A.~Bayramian, K.~D. Chesnut, A.~Erlandson, C.~W. Siders,
  E.~Sistrunk, T.~Spinka, and C.~Haefner.
\newblock {Scaling of petawatt-class lasers to multi-kHZ repetition rates}.
\newblock In Joachim Hein and Thomas~J. Butcher, editors, {\em High-Power,
  High-Energy, and High-Intensity Laser Technology IV}, volume 11033, pages
  1--8. International Society for Optics and Photonics, SPIE, 2019.

\bibitem{Siders19}
Craig~W. Siders, Thomas Galvin, Alvin Erlandson, Andrew Bayramian, Brendan
  Reagan, Emily Sistrunk, Thomas Spinka, and Constantin Haefner.
\newblock {Wavelength Scaling of Laser Wakefield Acceleration for the EuPRAXIA
  Design Point}.
\newblock {\em Instruments}, 3(3), 2019.

\bibitem{Workshop13}
{\em {Workshop on Laser Technology for Accelerators Summary Report}},
  Department of Energy, 2013.

\bibitem{Workshop17}
{Report of Workshop on Laser Technology for k-BELLA and Beyond}.
\newblock Technical report, LBNL, 2017.

\bibitem{falcone20}
Roger Falcone, Felicie Albert, Farhat Beg, Siegfried Glenzer, Todd Ditmire, Tom
  Spinka, and Jonathan Zuegel.
\newblock {{\it Workshop Report: Brightest Light Initiative}}.
\newblock {\em arXiv:2002.09712}, Optical Society of America, 2020.

\bibitem{DOE2016}
DOE.
\newblock Advanced accelerator development strategy report: {DOE} advanced
  accelerator concepts research roadmap workshop.
\newblock Technical report, February 2016.

\bibitem{benedetti.arxiv.2022b}
C.~Benedetti, S.~S. Bulanov, E.~Esarey, C.~G. R. Geddes A.~J. Gonsalves, P.~M.
  Jacobs, S.~Knapen, B.~Nachman, K.~Nakamura, S.~Pagan Griso, C.~B. Schroeder,
  D.~Terzani, J.~van Tilborg, M.~Turner, W.~M. Yao, R.~Bernstein, V.~Shiltsev,
  S.~J. Gessner, M.~J. Hogan, T.~Nelson, C.~Jing, I.~Low, X.~Lu, R.~Yoshida,
  C.~Lee, P.~Meade, N.~Vafaei-Najafabadi, P.~Muggli, P.~Musumeci, M.~Palmer,
  E.~Prebys, L.~Visinelli, C.~A. Aidala, and A.~G.~R. Thomas.
\newblock Whitepaper submitted to snowmass21: Advanced accelerator linear
  collider demonstration facility at intermediate energy, arxiv: 2203.08425,
  2022.

\bibitem{ANAR17}
Brigitte Cros and Patric Muggli, editors.
\newblock {\em Towards a Proposal for an Advanced Linear Collider, Report on
  the Advanced and Novel Accelerators for High Energy Physics Roadmap
  Workshop}, number CH-1211. CERN, 2017.

\bibitem{KirbyPAC07}
N.~Kirby, M.~Berry, I.~Blumenfeld, M.~J. Hogan, R.~Ischebeck, and R.~Siemann.
\newblock Emittance growth from multiple coulomb scattering in a plasma
  wakefield accelerator.
\newblock In {\em 2007 IEEE Particle Accelerator Conference (PAC)}, pages
  3097--3099, 2007.

\bibitem{zhao2020modeling}
Y~Zhao, R~Lehe, A~Myers, M~Th{\'e}venet, A~Huebl, CB~Schroeder, and J-L Vay.
\newblock Modeling of emittance growth due to coulomb collisions in
  plasma-based accelerators.
\newblock {\em Physics of Plasmas}, 27(11):113105, 2020.

\bibitem{Huang07}
C.~Huang, W.~Lu, M.~Zhou, C.~E. Clayton, C.~Joshi, W.~B. Mori, P.~Muggli,
  S.~Deng, E.~Oz, T.~Katsouleas, M.~J. Hogan, I.~Blumenfeld, F.~J. Decker,
  R.~Ischebeck, R.~H. Iverson, N.~A. Kirby, and D.~Walz.
\newblock Hosing instability in the blow-out regime for plasma-wakefield
  acceleration.
\newblock {\em Phys. Rev. Lett.}, 99(25):255001, Dec 2007.

\bibitem{mehrling.prl.2017}
T.~J. Mehrling, R.~A. Fonseca, A.~Martinez de~la Ossa, and J.~Vieira.
\newblock Mitigation of the hose instability in plasma-wakefield accelerators.
\newblock {\em Phys. Rev. Lett.}, 118:174801, Apr 2017.

\bibitem{WLu.prl.2006}
W.~Lu, C.~Huang, M.~Zhou, W.~B. Mori, and T.~Katsouleas.
\newblock Nonlinear theory for relativistic plasma wakefields in the blowout
  regime.
\newblock {\em Phys. Rev. Lett.}, 96:165002, Apr 2006.

\bibitem{XXu.prl.2016}
X.~L. Xu, J.~F. Hua, Y.~P. Wu, C.~J. Zhang, F.~Li, Y.~Wan, C.-H. Pai, W.~Lu,
  W.~An, P.~Yu, M.~J. Hogan, C.~Joshi, and W.~B. Mori.
\newblock Physics of phase space matching for staging plasma and traditional
  accelerator components using longitudinally tailored plasma profiles.
\newblock {\em Phys. Rev. Lett.}, 116:124801, Mar 2016.

\bibitem{ariniello.prab.2019}
R.~Ariniello, C.~E. Doss, K.~Hunt-Stone, J.~R. Cary, and M.~D. Litos.
\newblock Transverse beam dynamics in a plasma density ramp.
\newblock {\em Phys. Rev. Accel. Beams}, 22:041304, Apr 2019.

\bibitem{An17}
Weiming An, Wei Lu, Chengkun Huang, Xinlu Xu, Mark~J. Hogan, Chan Joshi, and
  Warren~B. Mori.
\newblock Ion motion induced emittance growth of matched electron beams in
  plasma wakefields.
\newblock {\em Phys. Rev. Lett.}, 118:244801, Jun 2017.

\bibitem{benedetti.prab.2017}
C.~Benedetti, C.~B. Schroeder, E.~Esarey, and W.~P. Leemans.
\newblock Emittance preservation in plasma-based accelerators with ion motion.
\newblock {\em Phys. Rev. Accel. Beams}, 20:111301, Nov 2017.

\bibitem{Mehrling2018}
T.{\hspace{0.167em}}J. Mehrling et~al.
\newblock Suppression of beam hosing in plasma accelerators with ion motion.
\newblock {\em Physical Review Letters}, 121(26), December 2018.

\bibitem{burov.arxiv.2018}
A.~Burov, S.~Nagaitsev, and V.~Lebedev.
\newblock Beam breakup mitigation by ion mobility in plasma acceleration, 2018.

\bibitem{Snowmass:ILC}
Alexander Aryshev, Ties Behnke, Mikael Berggren, James Brau, Nathaniel Craig,
  Ayres Freitas, Frank Gaede, Spencer Gessner, Stefania Gori, Christophe
  Grojean, et~al.
\newblock The international linear collider: Report to snowmass 2021.
\newblock {\em arXiv preprint arXiv:2203.07622}, 2022.

\bibitem{LindstromArxiv2021}
Carl~A. Lindstrøm.
\newblock Self-correcting longitudinal phase space in a multistage plasma
  accelerator, 2021.

\bibitem{muggli.awake.snowmass}
P.~Muggli and AWAKE Collaboration.
\newblock White paper: Awake, plasma wakefield acceleration of electron bunches
  for near and long term particle physics applications, 2022.

\bibitem{gai2012short}
W~Gai, JG~Power, and C~Jing.
\newblock Short-pulse dielectric two-beam acceleration.
\newblock {\em Journal of Plasma Physics}, 78(4):339--345, 2012.

\bibitem{Simakov_2022}
E.I. Simakov, G.~Andonian, S.S. Baturin, and P.~Manwani.
\newblock Limiting effects in drive bunch beam dynamics in beam-driven
  accelerators: instability and collective effects.
\newblock {\em Journal of Instrumentation}, 17(05):P05013, may 2022.

\bibitem{zholents:ipac2020-tuvir08}
A.~Zholents, S.~Baturin, D.S. Doran, W.G. Jansma, M.~Kasa, A.~Nassiri, P.~Piot,
  J.G. Power, A.E. Siy, S.~Sorsher, K.J. Suthar, W.H. Tan, E.~Trakhtenberg,
  G.J. Waldschmidt, and J.Z. Xu.
\newblock {A Compact High Repetition Rate Free-Electron Laser Based on the
  Advanced Wakefield Accelerator Technology}.
\newblock presented at IPAC'20 in Caen, France, unpublished, 06 2020.

\bibitem{vieira2014nonlinear}
Jorge Vieira and JT~Mendon{\c{c}}a.
\newblock Nonlinear laser driven donut wakefields for positron and electron
  acceleration.
\newblock {\em Phys. Rev. Lett.}, 112(21):215001, 2014.

\bibitem{zhou.e+.prl.2021}
Shiyu Zhou, Jianfei Hua, Weiming An, Warren~B. Mori, Chan Joshi, Jie Gao, and
  Wei Lu.
\newblock High efficiency uniform wakefield acceleration of a positron beam
  using stable asymmetric mode in a hollow channel plasma.
\newblock {\em Phys. Rev. Lett.}, 127:174801, Oct 2021.

\bibitem{Diederichs20}
S.~Diederichs, C.~Benedetti, E.~Esarey, J.~Osterhoff, and C.~B. Schroeder.
\newblock High-quality positron acceleration in beam-driven plasma
  accelerators.
\newblock {\em Phys. Rev. Accel. Beams}, 23:121301, Dec 2020.

\bibitem{silva2021stable}
T~Silva, LD~Amorim, MC~Downer, MJ~Hogan, V~Yakimenko, R~Zgadzaj, and J~Vieira.
\newblock Stable positron acceleration in thin, warm, hollow plasma channels.
\newblock {\em Phys. Rev. Lett.}, 127(10):104801, 2021.

\bibitem{1998hep.ex10019T}
Valery {Telnov}.
\newblock {Gamma-gamma, gamma-electron colliders}.
\newblock {\em arXiv e-prints}, pages hep--ex/9810019, October 1998.

\bibitem{Yakimenko2019}
V.~Yakimenko et~al.
\newblock Prospect of studying nonperturbative {QED} with beam-beam collisions.
\newblock {\em Physical Review Letters}, 122(19), May 2019.

\bibitem{peskin.july.2022}
M.~Peskin.
\newblock {Private Communication}, 2022.

\bibitem{Blondel98}
Blondel.
\newblock A muon collider as z factory, 1998.

\bibitem{EuroRoadmap2022}
European strategy for particle physics accelerator r\&d roadmap,
  "https://arxiv.org/ftp/arxiv/papers/2201/2201.07895.pdf".
\newblock {\em Report on the European Strategy For Particle Physics}, 2022.

\bibitem{Assman2020}
R.~W.~Assman {\it et~al.}
\newblock Eupraxia conceptual design report.
\newblock {\em Euro. Phys. J. Spec. Top.}, 229:3675, 2020.

\bibitem{AWAKE-Nature}
AWAKE Collaboration.
\newblock {Acceleration of electrons in the plasma wakefield of a proton
  bunch}.
\newblock {\em Nature}, 561(7723):363--367, 2018.

\bibitem{alegro.arxiv.2020}
B.~Cros and P.~Muggli.
\newblock {ALEGRO input for the 2020 update of the European Strategy}.
\newblock {\em arXiv:1901.08436 [physics.acc-ph]}, 1 2019.

\bibitem{BRNmanufacturing2020}
{Jenk,~Cynthia,~and~others}.
\newblock {Basic Research Needs for Transformative Manufacturing: Fundamental
  science to revolutionize manufacturing. Report from the US Department of
  Energy, Office of Basic Energy Sciences Workshop.}
\newblock 3 2020.

\bibitem{emma2022snowmass}
Claudio Emma, Jeroen van Tilborg, F{\'e}licie Albert, Luca Labate, Joel
  England, Spencer Gessner, Frederico Fiuza, Lieselotte Obst-Huebl, Alexander
  Zholents, Alex Murokh, et~al.
\newblock Snowmass 2021 accelerator frontier white paper: Near term
  applications driven by advanced accelerator concepts.
\newblock {\em arXiv preprint arXiv:2203.09094}, 2022.

\bibitem{near-term-Snowmass21}
Near term applications driven by advanced accelerator concepts.
\newblock 2022.
\newblock {Snowmass21} white paper, to be submitted to arXiv.

\bibitem{snowmass2013}
M~Bardeen, D~Cronin-Hennessy, RM~Barnett, P~Bhat, K~Cecire, K~Cranmer,
  T~Jordan, I~Karliner, J~Lykken, P~Norris, et~al.
\newblock Planning the future of us particle physics (snowmass 2013): Chapter
  10: Communication, education, and outreach.
\newblock {\em arXiv preprint arXiv:1401.6119}, 2014.

\bibitem{nanni:ccc}
Emilio~A Nanni, Martin Breidenbach, Caterina Vernieri, Sergey Belomestnykh,
  Pushpalatha Bhat, Sergei Nagaitsev, Mei Bai, Tim Barklow, Ankur Dhar, Ram~C
  Dhuley, et~al.
\newblock C3 demonstration research and development plan.
\newblock {\em arXiv preprint arXiv:2203.09076}, 2022.

\bibitem{Emma2021}
C.~Emma, J.~van Tilborg, R.~Assmann, S.~Barber, A.~Cianchi, S.~Corde, M.~E.
  Couprie, R.~D'Arcy, M.~Ferrario, A.~F. Habib, B.~Hidding, M.~J. Hogan, C.~B.
  Schroeder, A.~Marinelli, M.~Labat, R.~Li, J.~Liu, A.~Loulergue, J.~Osterhoff,
  A.~R. Maier, B.~W.~J. McNeil, , and W.~Wang.
\newblock Free electron lasers driven by plasma accelerators: status and
  near-term prospects.
\newblock {\em {High Power Laser Sci. Eng.}}, {9}:e57, {2021}.

\bibitem{Zholents2020}
A.~Zholents et~al.
\newblock A compact high repetition rate free-electron laser based on the
  advanced wakefield accelerator technology.
\newblock In {\em IPAC20}. JACoW, May 2020.

\bibitem{Rosenzweig2020}
J.~B. Rosenzweig, N.~Majernik, R.~R. Robles, G.~Andonian, O.~Camacho,
  A.~Fukasawa, A.~Kogar, G.~Lawler, Jianwei Miao, P.~Musumeci, B.~Naranjo,
  Y.~Sakai, R.~Candler, B.~Pound, C.~Pellegrini, C.~Emma, A.~Halavanau,
  J.~Hastings, Z.~Li, M.~Nasr, S.~Tantawi, P.~Anisimov, B.~Carlsten,
  F.~Krawczyk, E.~Simakov, L.~Faillace, M.~Ferrario, B.~Spataro, S.~Karkare,
  J.~Maxson, Y.~Ma, J.~Wurtele, A.~Murokh, A.~Zholents, A.~Cianchi, D.~Cocco, ,
  and S.~B. van~der Geer.
\newblock An ultra-compact x-ray free-electron laser.
\newblock {\em {New J. Phys.}}, {22}:093067, {2020}.

\bibitem{Corde_RMP_2013}
S.~Corde, K.~Ta~Phuoc, G.~Lambert, R.~Fitour, V.~Malka, A.~Rousse, A.~Beck, and
  E.~Lefebvre.
\newblock Femtosecond x rays from laser-plasma accelerators.
\newblock {\em Rev. Mod. Phys.}, 85:1--48, Jan 2013.

\bibitem{Albert_PPCF_2016}
Felicie Albert and Alec G.~R. Thomas.
\newblock Applications of laser wakefield accelerator-based light sources.
\newblock {\em Plasma Phys. Control. Fusion}, 58(10):103001, 2016.

\bibitem{Glinec_PRL_2005}
Y.~Glinec, J.~Faure, L.~Le Dain, S.~Darbon, T.~Hosokai, J.~J. Santos,
  E.~Lefebvre, J.~P. Rousseau, F.~Burgy, B.~Mercier, and V.~Malka.
\newblock High-resolution $\ensuremath{\gamma}$-ray radiography produced by a
  laser-plasma driven electron source.
\newblock {\em Phys. Rev. Lett.}, 94:025003, Jan 2005.

\bibitem{Benismail_NIMA_2011}
A.~Ben-Ismaïl, J.~Faure, and V.~Malka.
\newblock Optimization of gamma-ray beams produced by a laser-plasma
  accelerator.
\newblock {\em Nucl. Instrum. Methods Phys. Res. A}, 629(1):382--386, 2011.

\bibitem{Gadjev2019}
I.~Gadjev, N.~Sudar, M.~Babzien, J.~Duris, P.~Hoang, M.~Fedurin, K.~Kusche,
  R.~Malone, P.~Musumeci, M.~Palmer, I.~Pogorelsky, M.~Polyanskiy, Y.~Sakai,
  C.~Swinson, O.~Williams, and J.~B. Rosenzweig.
\newblock An inverse free electron laser acceleration-driven compton scattering
  x-ray source.
\newblock {\em Scientific Reports}, 9(1), January 2019.

\bibitem{Sudar2020}
N.~Sudar, P.~Musumeci, A.~Ovodenko, A.~Murokh, M.~Polyanskiy, I.~Pogorelsky,
  M.~Fedurin, C.~Swinson, K.~Kusche, M.~Babzien, and M.~Palmer.
\newblock Burst mode mhz repetition rate inverse free electron laser
  acceleration.
\newblock {\em Phys. Rev. Accel. Beams}, 23:051301, May 2020.

\bibitem{Svendsen2021}
Kristoffer Svendsen, Diego Gu{\'{e}}not, Jonas~Bj\"{o}rklund Svensson,
  Kristoffer Petersson, Anders Persson, and Olle Lundh.
\newblock A focused very high energy electron beam for fractionated
  stereotactic radiotherapy.
\newblock {\em Scientific Reports}, 11(1), March 2021.

\bibitem{Labate2020}
Luca Labate, Daniele Palla, Daniele Panetta, Federico Avella, Federica Baffigi,
  Fernando Brandi, Fabio~Di Martino, Lorenzo Fulgentini, Antonio Giulietti,
  Petra K\"{o}ster, Davide Terzani, Paolo Tomassini, Claudio Traino, and
  Leonida~A. Gizzi.
\newblock Toward an effective use of laser-driven very high energy electrons
  for radiotherapy: Feasibility assessment of multi-field and intensity
  modulation irradiation schemes.
\newblock {\em Scientific Reports}, 10(1), October 2020.

\bibitem{Kokurewicz2019}
K.~Kokurewicz, E.~Brunetti, G.~H. Welsh, S.~M. Wiggins, M.~Boyd, A.~Sorensen,
  A.~J. Chalmers, G.~Schettino, A.~Subiel, C.~DesRosiers, and D.~A.
  Jaroszynski.
\newblock Focused very high-energy electron beams as a novel radiotherapy
  modality for producing high-dose volumetric elements.
\newblock {\em Scientific Reports}, 9(1), July 2019.

\bibitem{Linz2016}
Ute Linz and Jose Alonso.
\newblock {Laser-driven ion accelerators for tumor therapy revisited}.
\newblock {\em Phys. Rev. Accel. Beams}, 19(12):124802, 2016.

\bibitem{gessner2020beamdump}
S~Gessner et~al.
\newblock Beamdump experiments driven by a plasma wakefield accelerator.
\newblock {\em Snowmass Letter of Intent}, 2020.

\bibitem{BeamTestFacilities_Snowmass2021}
John Power, Christine Clarke, Michael Downer, Eric Esarey, Cameron Geddes,
  Mark~J. Hogan, Georg~Heinz Hoffstaetter, Chunguang Jing, Sergei Nagaitsev,
  Mark Palmer, Philippe Piot, Carl Schroeder, Donald Umstadter, Navid
  Vafaei-Najafabadi, Alexander Valishev, Louise Willingale, and Vitaly
  Yakimenko.
\newblock Beam test facilities for r\&d in accelerator science and
  technologies, 2022.

\bibitem{hepap-meeting-2019}
Update on p5 gard subpanel recommendations, 2019.

\bibitem{Doche2017}
A.~Doche, C.~Beekman, S.~Corde, J.~M. Allen, C.~I. Clarke, J.~Frederico, S.~J.
  Gessner, S.~Z. Green, M.~J. Hogan, B.~O'Shea, V.~Yakimenko, W.~An, C.~E.
  Clayton, C.~Joshi, K.~A. Marsh, W.~B. Mori, N.~Vafaei-Najafabadi, M.~D.
  Litos, E.~Adli, C.~A. Lindstr{\o}m, and W.~Lu.
\newblock {Acceleration of a trailing positron bunch in a plasma wakefield
  accelerator}.
\newblock {\em Scientific Reports}, 7(1):1--7, 2017.

\bibitem{palmer-2011}
Charlotte A.~J. Palmer, N.~P. Dover, I.~Pogorelsky, M.~Babzien, G.~I.
  Dudnikova, M.~Ispiriyan, M.~N. Polyanskiy, J.~Schreiber, P.~Shkolnikov,
  V.~Yakimenko, and Z.~Najmudin.
\newblock Monoenergetic proton beams accelerated by a radiation pressure driven
  shock.
\newblock {\em Phys. Rev. Lett.}, 106:014801, Jan 2011.

\bibitem{babzien-2006}
Marcus Babzien, Ilan Ben-Zvi, Karl Kusche, Igor~V. Pavlishin, Igor~V.
  Pogorelsky, David~P. Siddons, Vitaly Yakimenko, David Cline, Feng Zhou,
  Tachishige Hirose, Yoshio Kamiya, Tetsuro Kumita, Tsunehiko Omori, Junji
  Urakawa, and Kaoru Yokoya.
\newblock Observation of the second harmonic in thomson scattering from
  relativistic electrons.
\newblock {\em Phys. Rev. Lett.}, 96:054802, Feb 2006.

\bibitem{yu-2000}
L.-H Yu, M~Babzien, I~Ben-Zvi, L.F DiMauro, A~Doyuran, W~Graves, E~Johnson,
  S~Krinsky, R~Malone, I~Pogorelsky, J~Skaritka, G~Rakowsky, L~Solomon, X.J
  Wang, M~Woodle, V~Yakimenko, S.G Biedron, J.N Galayda, E~Gluskin, J~Jagger,
  V~Sajaev, and I~Vasserman.
\newblock First lasing of a high-gain harmonic generation free- electron laser
  experiment.
\newblock {\em Nuclear Instruments and Methods in Physics Research Section A:
  Accelerators, Spectrometers, Detectors and Associated Equipment},
  445(1):301--306, 2000.

\bibitem{hoang-2018}
P.~D. Hoang, G.~Andonian, I.~Gadjev, B.~Naranjo, Y.~Sakai, N.~Sudar,
  O.~Williams, M.~Fedurin, K.~Kusche, C.~Swinson, P.~Zhang, and J.~B.
  Rosenzweig.
\newblock Experimental characterization of electron-beam-driven wakefield modes
  in a dielectric-woodpile cartesian symmetric structure.
\newblock {\em Phys. Rev. Lett.}, 120:164801, Apr 2018.

\bibitem{picard22}
Julian Picard, Ivan Mastovsky, Michael~A. Shapiro, Richard~J. Temkin, Xueying
  Lu, Manoel Conde, D.~Scott Doran, Gwanghui Ha, John~G. Power, Jiahang Shao,
  Eric~E. Wisniewski, and Chunguang Jing.
\newblock Generation of 565 mw of $x$-band power using a metamaterial power
  extractor for structure-based wakefield acceleration.
\newblock {\em Phys. Rev. Accel. Beams}, 25:051301, May 2022.

\bibitem{fruhling-2021}
Colton Fruhling, Grigory Golovin, and Donald Umstadter.
\newblock Attosecond electron bunch measurement with coherent nonlinear thomson
  scattering.
\newblock {\em Phys. Rev. Accel. Beams}, 23:072802, Jul 2020.

\bibitem{golovin2018electron}
Grigory Golovin, Wenchao Yan, Ji~Luo, Colton Fruhling, Dan Haden, Baozhen Zhao,
  Cheng Liu, Min Chen, Shouyuan Chen, Ping Zhang, et~al.
\newblock Electron trapping from interactions between laser-driven relativistic
  plasma waves.
\newblock {\em Physical review letters}, 121(10):104801, 2018.

\bibitem{england2022laser}
RJ~England, D~Filippetto, G~Torrisi, A~Bacci, G~Della~Valle, D~Mascali,
  GS~Mauro, G~Sorbello, P~Musumeci, J~Scheuer, et~al.
\newblock Laser-driven structure-based accelerators.
\newblock {\em arXiv preprint arXiv:2203.08981}, 2022.

\bibitem{ariniello2022channeling}
Robert Ariniello, Sebastien Corde, Xavier Davoine, Henrik Ekerfelt, Frederico
  Fiuza, Max Gilljohann, Laurent Gremillet, Yuliia Mankovska, Henryk Piekarz,
  Pablo San~Miguel Claveria, et~al.
\newblock Channeling acceleration in crystals and nanostructures and studies of
  solid plasmas: New opportunities.
\newblock {\em arXiv preprint arXiv:2203.07459}, 2022.

\bibitem{sahai:disambiguation}
Aakash~A Sahai.
\newblock Nanomaterials based nanoplasmonic accelerators and light-sources
  driven by particle-beams.
\newblock {\em IEEE Access}, 9:54831--54839, 2021.

\end{thebibliography}

\end{document}